\documentclass[longauth]{aa}

\usepackage{natbib}
\usepackage[breaklinks=true]{hyperref} 
\bibpunct{(}{)}{;}{a}{}{,} 

\usepackage{graphicx}
\usepackage{txfonts}
\usepackage{textcomp}
\usepackage{url}
\hypersetup{colorlinks=true,citecolor=blue}

\def\ltsim{\hbox{\raise 2pt \hbox {$<$} \kern-1.1em \lower 4pt \hbox {$\sim$}}}
\def\ltapprox{\hbox{\raise 2pt \hbox {$<$} \kern-1.1em \lower 5pt \hbox
{$\approx$}}}
\def\gtsim{\hbox{\raise 2pt \hbox {$>$} \kern-1.1em \lower 4pt \hbox {$\sim$}}}
\def\gtapprox{\hbox{\raise 2pt \hbox {$>$} \kern-1.1em \lower 5pt \hbox
{$\approx$}}}

\def\arcsec{$^{\rm \prime\prime}$}
\def\arcmin{$^{\rm \prime}$}
\def\degrees{$^{\rm \circ}$}

\begin{document}

   \title{Wide-field LOFAR imaging of the field around the double-double radio galaxy
     B1834+620:} 
   \subtitle{A fresh view on a restarted AGN and doubeltjes.}

   \author{E.~Orr{\`u}\inst{1,2}\and 
S.~van Velzen\inst{2,3}\and 
R. F.~Pizzo\inst{1}\and 
S.~Yatawatta\inst{1,4}\and 
R.~Paladino\inst{5,6} \and 
M.~Iacobelli\inst{1} \and
M.~Murgia\inst{7} \and 
H.~Falcke\inst{1,2,8} \and
R.~Morganti\inst{1,4} \and 
A. G.~de Bruyn\inst{1,4} \and 
C.~Ferrari\inst{9} \and
J.~Anderson\inst{10}\and 
A.~Bonafede\inst{11}\and
D.~Mulcahy\inst{12}\and
A.~Asgekar\inst{1,13}\and 
I.~M.~Avruch\inst{14,4}\and 
R.~Beck\inst{8}\and 
M.~E.~Bell\inst{15}\and 
I.~van Bemmel\inst{1,16}\and 
M.~J.~Bentum\inst{1,17}\and 
G.~Bernardi\inst{18}\and 
P.~Best\inst{19}\and 
F.~Breitling\inst{20}\and 
J.~W.~Broderick\inst{21}\and 
M.~Br\"uggen\inst{11}\and 
H.~R.~Butcher\inst{22}\and 
B.~Ciardi\inst{23}\and 
J.~E.~Conway\inst{24}\and 
A.~Corstanje\inst{2}\and 
E.~de Geus\inst{1,25}\and 
A.~Deller\inst{1}\and 
S.~Duscha\inst{1}\and 
J.~Eisl\"offel\inst{26}\and 
D.~Engels\inst{27}\and 
W.~Frieswijk\inst{1}\and 
M.~A.~Garrett\inst{1,28}\and 
J.~Grie\ss{}meier\inst{29,30}\and 
A.~W.~Gunst\inst{1}\and 
J.~P.~Hamaker\inst{1}\and
G.~Heald\inst{1}\and 
M.~Hoeft\inst{26}\and 
A.J. ~van der Horst\inst{31}\and 
H.~Intema\inst{28,32}\and 
E.~Juette\inst{33}\and 
J.~Kohler\inst{8}\and 
V.~I.~Kondratiev\inst{2,34}\and 
M.~Kuniyoshi\inst{35}\and 
G.~Kuper\inst{1}\and 
M.~Loose\inst{1}\and 
P.~Maat\inst{1}\and 
G.~Mann\inst{20}\and 
S.~Markoff\inst{36}\and 
R. McFadden\inst{1}\and 
D.~McKay-Bukowski\inst{37,38}\and 
G.~Miley\inst{28}\and 
J..~Moldon\inst{1}\and 
G.~Molenaar\inst{36}\and 
H.~Munk\inst{1}\and 
A.~Nelles\inst{2}\and 
H.~Paas\inst{39}\and 
M.~Pandey-Pommier\inst{40}\and 
V.~N.~Pandey\inst{1}\and 
G.~Pietka\inst{41}\and 
A.~G.~Polatidis\inst{1}\and 
W.~Reich\inst{8}\and 
H.~R\"ottgering\inst{28}\and 
A.~Rowlinson\inst{15}\and 
A.~Scaife\inst{12}\and
A.~Schoenmakers\inst{1}\and 
D.~Schwarz\inst{42}\and 
M.~Serylak\inst{41}\and 
A.~Shulevski\inst{1,4}\and 
O.~Smirnov\inst{43,44}\and 
M.~Steinmetz\inst{20}\and 
A.~Stewart\inst{41}\and 
J.~Swinbank\inst{36}\and 
M.~Tagger\inst{29}\and 
C.~Tasse\inst{45}\and 
S.~Thoudam\inst{2}\and 
M.~C.~Toribio\inst{1}\and 
R.~Vermeulen\inst{1}\and 
C. Vocks\inst{20}\and 
R.~J.~van Weeren\inst{18}\and 
R.~A.~M.~J.~Wijers\inst{36}\and 
M.~W.~Wise\inst{1,36}\and 
O.~Wucknitz\inst{8}
}
 
\offprints{E. Orr{\`u} (orru@astron.nl)}

\institute{
ASTRON, the Netherlands Institute for Radio Astronomy, Postbus 2, 7990 AA, Dwingeloo, The Netherlands 
\and Department of Astrophysics/IMAPP, Radboud University, P.O. Box 9010, 6500 GL Nijmegen, The Netherlands 
\and Center for Astrophysical Sciences, Department of Physics \& Astronomy, The Johns Hopkins University, Baltimore, Maryland 21218, USA 
\and Kapteyn Astronomical Institute, PO Box 800, 9700 AV Groningen, The Netherlands 
\and Department of Physics and Astronomy, University of Bologna, via Berti Pichat 6/2, I-40127 Bologna, Italy 
\and INAF - Osservatorio di Radioastronomia, Via P. Gobetti, 101 40129, Bologna 
\and INAF-Osservatorio Astronomico di Cagliari, Via della Scienza 5, I-09047, Selargius (Cagliari), Italy 
\and Max-Planck-Institut f\"{u}r Radioastronomie, Auf dem H\"ugel 69, 53121 Bonn, Germany 
\and Laboratoire Lagrange, UMR 7293, Universit\'e de Nice Sophia Antipolis, CNRS, Observatoire de la C\^ote d'Azur, 06300, Nice, France 
\and Helmholtz-Zentrum Potsdam, DeutschesGeoForschungsZentrum GFZ, Department 1: Geodesy and Remote Sensing, Telegrafenberg, A17, 14473 Potsdam, Germany 
\and University of Hamburg, Gojenbergsweg 112, 21029 Hamburg, Germany 
\and Jodrell Bank Centre for Astrophysics, School of Physics and Astronomy, The University of Manchester, Oxford Road, Manchester M13 9PL, UK 
\and Shell Technology Center, Bangalore, India 
\and SRON Netherlands Insitute for Space Research, PO Box 800, 9700 AV Groningen, The Netherlands 
\and CSIRO Australia Telescope National Facility, PO Box 76, Epping NSW 1710, Australia 
\and Joint Institute for VLBI in Europe, Dwingeloo, Postbus 2, 7990 AA The Netherlands 
\and University of Twente, The Netherlands 
\and Harvard-Smithsonian Center for Astrophysics, 60 Garden Street, Cambridge, MA 02138, USA 
\and Institute for Astronomy, University of Edinburgh, Royal Observatory of Edinburgh, Blackford Hill, Edinburgh EH9 3HJ, UK 
\and Leibniz-Institut f\"{u}r Astrophysik Potsdam (AIP), An der Sternwarte 16, 14482 Potsdam, Germany 
\and School of Physics and Astronomy, University of Southampton, Southampton, SO17 1BJ, UK 
\and Research School of Astronomy and Astrophysics, Australian National University, Mt Stromlo Obs., via Cotter Road, Weston, A.C.T. 2611, Australia 
\and Max Planck Institute for Astrophysics, Karl Schwarzschild Str. 1, 85741 Garching, Germany 
\and Onsala Space Observatory, Dept. of Earth and Space Sciences, Chalmers University of Technology, SE-43992 Onsala, Sweden 
\and SmarterVision BV, Oostersingel 5, 9401 JX Assen 
\and Th\"{u}ringer Landessternwarte, Sternwarte 5, D-07778 Tautenburg, Germany 
\and Hamburger Sternwarte, Gojenbergsweg 112, D-21029 Hamburg 
\and Leiden Observatory, Leiden University, PO Box 9513, 2300 RA Leiden, The Netherlands 
\and LPC2E - Universite d'Orleans/CNRS 
\and Station de Radioastronomie de Nancay, Observatoire de Paris - CNRS/INSU, USR 704 - Univ. Orleans, OSUC , route de Souesmes, 18330 Nancay, France 
\and Department of Physics, The George Washington University, 725 21st Street NW, Washington, DC 20052, USA 
\and National Radio Astronomy Observatory, 520 Edgemont Road, Charlottesville, VA 22903-2475, USA 
\and Astronomisches Institut der Ruhr-Universit\"{a}t Bochum, Universitaetsstrasse 150, 44780 Bochum, Germany 
\and Astro Space Center of the Lebedev Physical Institute, Profsoyuznaya str. 84/32, Moscow 117997, Russia 
\and National Astronomical Observatory of Japan, Japan 
\and Anton Pannekoek Institute, University of Amsterdam, Postbus 94249, 1090 GE Amsterdam, The Netherlands 
\and Sodankyl\"{a} Geophysical Observatory, University of Oulu, T\"{a}htel\"{a}ntie 62, 99600 Sodankyl\"{a}, Finland 
\and STFC Rutherford Appleton Laboratory,  Harwell Science and Innovation Campus,  Didcot  OX11 0QX, UK 
\and Center for Information Technology (CIT), University of Groningen, The Netherlands 
\and Centre de Recherche Astrophysique de Lyon, Observatoire de Lyon, 9 av Charles Andr\'{e}, 69561 Saint Genis Laval Cedex, France 
\and Astrophysics, University of Oxford, Denys Wilkinson Building, Keble Road, Oxford OX1 3RH 
\and Fakult\"{a}t f\"{u}r Physik, Universit\"{a}t Bielefeld, Postfach 100131, D-33501, Bielefeld, Germany 
\and Department of Physics and Elelctronics, Rhodes University, PO Box 94, Grahamstown 6140, South Africa 
\and SKA South Africa, 3rd Floor, The Park, Park Road, Pinelands, 7405, South Africa 
\and LESIA, UMR CNRS 8109, Observatoire de Paris, 92195   Meudon, France }

\authorrunning{Orr{\`u}, van Velzen, Pizzo, et al.}
\titlerunning{LOFAR observations of the double-double radio galaxy B1834+620}
   \date{Received; accepted}
   
\abstract
   {The existence of double-double radio galaxies (DDRGs) is 
    evidence for recurrent jet activity in AGN, as expected from standard accretion
    models. A detailed study of these rare sources provides new perspectives for investigating the AGN
    duty cycle, AGN-galaxy feedback, and accretion mechanisms. Large catalogues of radio sources, 
    on the other hand, provide statistical information about the evolution 
    of the radio-loud AGN population out to high redshifts.}
   {Using wide-field imaging with the LOFAR telescope, we  study both a well-known DDRG 
   as well as a large number of radio sources in the field of view.
   }
   {We present a high resolution image of the DDRG B1834+620
    obtained at 144\,MHz using LOFAR commissioning data.  Our image
    covers about 100 square degrees and contains
    over 1000 sources. 
     }
   {The four components of the DDRG B1834+620 have been resolved for
     the first time at 144\,MHz. Inner lobes were found to point towards the direction of the outer lobes, unlike standard FR~II
    sources. Polarized emission was detected at +60 rad
    m$^{-2}$ in the northern outer lobe. 
    The high spatial resolution allows the identification of a 
     large number of small double-lobed radio sources; roughly 10\%
    of all sources in the field are doubles with a separation smaller than 1\arcmin.
    }
   {The spectral fit of the four components is consistent with a
    scenario in which the outer lobes are still active or the jets
    recently switched off, while emission of the inner lobes is the
    result of a mix-up of new and old jet activity. From the presence of the newly extended features in the inner lobes of
    the DDRG, we can infer that the mechanism responsible for their
    formation is the bow shock that is driven by the newly launched
    jet.  
    We find that the density of the small doubles exceeds
    the density of FR~II sources with similar properties at
    1.4\,GHz, but this difference becomes smaller for low flux densities. 
    Finally, we show that the significant challenges of wide-field imaging
    (e.g., time and frequency variation of the beam, directional
    dependent calibration errors) can be solved using LOFAR commissioning data, thus demonstrating
    the potential of the full LOFAR telescope to discover millions of powerful AGN at redshift $z\sim1$.
    }


\maketitle
%

\section{Introduction}

The dichotomy that separates the active galactic nuclei (AGN) population into 
radio-loud and radio- quiet AGN has been known for a long time \citep{Kellermann89}. 
Radio-loud sources represent about 10$\%$ of the total AGN population, suggesting that either the radio-loud phase is active only for a fraction of the AGN
lifetime or that special physical properties of the black hole (BH) are required to
trigger the jet formation. Studies of the energetics of radio galaxies and the AGN's duty cycle can be used to
quantify the influence of jets in the environment. This information
can be used to develop evolution models \citep[see e.g.,][]{DiMatteo2005, Fabian2006b,
McNamara2007, Holt2008, McNamara2012, Morganti2013}. 

Jet intermittence is well supported by theory. According to the fundamental
plane of black hole accretion, which unifies non-thermal emission of
black holes over eight orders of magnitudes \citep*{Merloni03,Falcke04}, an analogy between AGN
and X-ray binaries can be inferred. Therefore, as observed for X-ray
binaries, AGN are expected to develop intermittent jets that correspond
to different accretion states
\citep{Fender2004,Koerding2008}. \cite{Best2009} found that 25\% of high mass galaxies (0.1 $<z<$ 0.3) host radio sources, although the radio lifetime is much less than the cosmological one.
Observations of double-double radio galaxies (DDRGs) provide one of the
best confirmations of the recurrence of the jet activity
\citep[e.g.][]{Schoenmakers2000,Konar2012}.

DDRGs are often found in samples of giant radio sources. \cite{Konar2004} reported the higher incidence of steep spectrum core among giant radio sources. These steep spectrum cores might be the new doubles whose sizes are too small to be resolved by their observations. Doubles are defined as a pair of double-lobed radio
sources aligned along the same axis, with a coinciding radio core \citep[][]{Schoenmakers2000}.
An important connection between the restarted jet activity and the
AGN-galaxy feedback is clearly confirmed by the detection of molecular
gas for the inner doubles of four DDRGs  \citep[][]{Schilizzi2001, Morganti2005, Saikia2009, Chandola2010, Labiano2013}.

The new International LOFAR Telescope \citep{LOFAR2013} can provide new insights into
the science of restarted AGNs. High-resolution, low-frequency radio images and spectral
index distribution maps simultaneously offer information about past and current nuclear activity. 
In this paper we make the first attempt to observe the
DDRG B1864+620 with the LOFAR . This source is a giant radio galaxy at z=0.5194, extending over
230\arcsec\  or 1.4 Mpc projected on the plane of the sky\footnote{Throughout we adopt
  H$_0$\,=\,71\,km\,s$^{-1}$\,Mpc$^{-1}$, $\Omega_m$\,=\,0.27,
  $\Omega_{\Lambda}$\,=\,0.73 \cite[]{Spergel2003}.}, and which has
been studied at several frequencies \citep{Schoenmakers2000a, Konar2012}.

A single LOFAR observation covers a large area of the sky ($\sim 10$ deg$^2$ at 150~MHz), 
enabling us to do ancillary science with our
data. In particular, the frequency range of our observation, centered
at 138.9\,MHz, implies that we predominantly detect sources with a steep spectral index, $\alpha<0$, $
S_\nu\propto \nu^\alpha$. The lobes of FR~II radio galaxies and
quasars \citep{Fanaroff1974} are a well-known population of steep
spectrum sources. 
The integrated flux density of most powerful sources in this
class, such as Cygnus~A, is $\sim 10$~mJy at $z=10$. Most
importantly, given their typical lobe separation of roughly 100 kpc, 
the expected angular separation of FR~IIs ranges from 5 to 30\arcsec\ at  $z>0.5$. As a result,  the high
resolution reached by LOFAR will allow us to unlock a large population
of powerful jets at high redshift by using only their morphological properties. 

The rest of this paper is organized as follows. 
The observations and calibration processes are discussed in Section 2; results
are discussed in Section 3;  in Section 4 we draw our
conclusions.

\section{Data} 

\subsection{Observation}
In this paper we present the results of an observation performed
during the LOFAR commissioning phase. 
The observation was performed using the LOFAR high-band antennas (HBA)
in the dual configuration, pointing towards the direction of the DDRG B1834+620 (RA: 18:35:10.92 DEC:
+62:04:08.10 J2000).  The observation was obtained with the full LOFAR
array (as available at that time), which  comprised 18 core stations in dual mode (CS), seven
remote stations (RS) and two international stations (IS). The data from
each HBA field in a core station were correlated as separate stations
in the LOFAR Blue Gene Correlator \citep{CEP}. In total, 45
stations were available \citep[][]{LOFAR2013}. The data
imaged and discussed in this paper include Dutch
stations. IS were excluded, as the data quality was lower and the UV coverage with only two stations was insufficient for adequate calibration. 

The target was observed from 21 May 2011 at
22:00 (UTC)  until 22 May 2011 at 05:00 (UTC) for a
total of seven hours. The instrument filter used was 110-190\,MHz, the
central frequency was set to 138.9\,MHz with a total bandwidth of 47.66\,MHz. 
The total bandwidth was divided into 244 sub-bands (SBs): each SB has
a bandwidth of 195.31\,kHz and was further divided into 64
channels of 3.05\,kHz bandwidth. Data were recorded with 2~s integration time.

\subsection{Data processing}
Besides detecting low brightness emission from the lobes of B1834+620
and polarized emission, the aim of this project is to produce a
wide-field image  to characterize faint  sources at low frequencies. 
Low-frequency observations are heavily corrupted by strong radio
frequency interference (RFI),
ionospheric effects, and strong
off-axis sources. To understand the importance of the latter, one has
to keep in mind that LOFAR imaging is all-sky imaging.
Via the side lobes of the HBA station beam, the so-called
A-team\footnote{The A-team (Cygnus~A, Cassiopeia~A, Virgo~A,
  Hercules~A, and Taurus~A) dominates the sky at low-frequency.} sources at tens of degrees from the phase center  significantly
influence the target field. Furthermore, delays in the clock synchronization between
LOFAR stations and the variability of the beam shape in time and frequency make the calibration process more
complex than standard interferometry, requiring the use of specialized software packages.  
In the following three sections we describe the calibration process applied to the
data  to overcome these challenges. 

\subsubsection{Pre-processing} 
The full resolution data-set was processed through the New Default
Pre-Processing Pipeline (NDPPP) to flag the baseline autocorrelations. 
Narrow-band RFI was flagged using the automatic {\it AOflagger} algorithm
\citep{Offringa2012, Offringa2010}.  
Ten stations were fully flagged owing to sensitivity loss or bad RFI.
The sensitivity loss for these stations was related to a problem with
the clock synchronization within a station and between
its elements, resulting in a 5~ns delay error which
affected the station sensitivity and calibration. (This problem has
been solved with the installation of new clock boards.) 
The total number of stations was reduced to 33, since ten stations were removed because of low sensitivity, and 
two IS were removed because of  poor UV-coverage.
After flagging, visual inspection of the data showed a big amplitude
bump in the visibilities during the second half of the observation,
which was unrelated to residual RFI. This excess was due to the effect of emission from Cassiopeia~A and Cygnus~A, 
located at 34\degrees\ and 25\degrees\ from the phase center,\ respectively. To properly subtract
the emission of these sources from the visibilities, the demixing algorithm was
applied to the full resolution data set \citep{Bas2007}. The output product of the demix
step resulted in a data set that was averaged to a frequency resolution of one SB and
a time resolution of ten seconds. Inspection of the data showed a
successful subtraction of the A-team in the case of 170 SBs, which
is considered in the following steps. Another run of NDPPP, using
the AOflagger procedure, was applied to remove low-level RFI that emerged after the demixing step. 

\begin{figure}[]
\includegraphics[width=9cm]{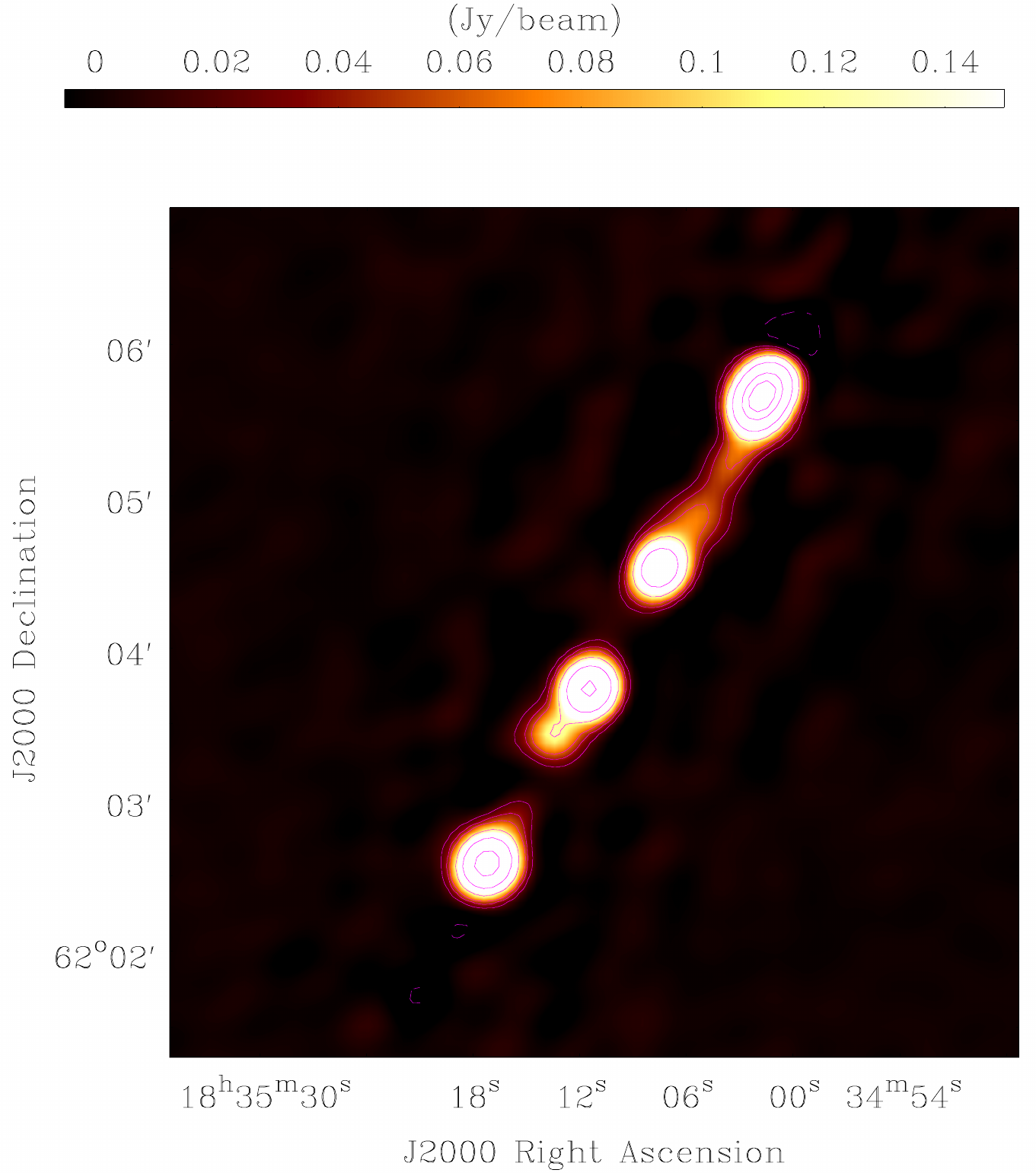}
\caption[]{Color scale shows the intensity of the LOFAR image at 144
  MHz, contours are 0.02, 0.04, 0.06, 0.12, 0.24, 0.5, and 0.7 Jy/beam
  (resolution 19\arcsec $\times$ 18\arcsec).}
\label{B1834+620}
\end{figure} 

\subsubsection{Calibration and imaging}
\label{calibration}
The observation was self-calibrated using BBS (Black Board
Selfcal). To this end, an initial model, including about 100 sources, was obtained by
analysing a Westerbork Synthesis Radio Telescope (WSRT) image at
139\,MHz of the field of B1834+620 through the source extraction
module Python Blob Detection and Source Measurement 
Software (PyBDSM)\footnote{PyBDSM is developed by D. Rafferty and N. Mohan,
  see the  LOFAR Imaging Cookbook, 
\newline
  \href{http://www.astron.nl/radio-observatory/lofar/lofar-imaging-cookbook}
  {http://www.astron.nl/radio-observatory/lofar/lofar-imaging-cookbook},
  and this link \href
  {http://home.strw.leidenuniv.nl/\~mohan/anaamika}{http://home.strw.leidenuniv.nl/\texttildelow
    mohan/anaamika}. The Astrophysics Source Code Library can be found in \href{http://adsabs.harvard.edu/abs/2015ascl.soft02007M}{http://adsabs.harvard.edu/abs/2015ascl.soft02007M}.} (see the end of this section for a
more detailed explanation). 
The calibration strategy consisted of solving for the four complex
terms of the G-Jones matrix (antenna gains) in the direction of the
phase center. During the solving process the theoretical beam
  model is used by BBS to predict the instrumental polarization. This
  prediction is limited by the accuracy of the beam model. Finally the calculated
gain solutions were applied to the data towards the
direction of the phase center for each time interval, without
  applying any smoothing or interpolation. 
 After the calibration, visibility data above a level of 30 Jy were clipped, since low signal-to-noise (S/N) gain solutions that were applied to the data
generated RFI-like spikes in the visibilities.  
Separate images were produced for each SB by using the AWimager imaging algorithm \citep{Tasse2013}. This imager applies the
aw-projection algorithm to simultaneously take into account the time
and frequency variability of the primary beam across the field-of-view
during synthesis observations and to remove the effects
of non-coplanar baselines when imaging a large field-of-view. 
The image resulting from the stacking of 170 images (rms\,=\,1.2 mJy/beam and resolution\,=\,24\arcsec $\times$ 21\arcsec) present two types of
artifacts: 1) Bright sources, from a few hundred milliJanskys up to 1\,Jy level
of flux density, in offset with respect to the phase center, showed radial spike-like features centered on the
source itself. In this case fast ionospheric phase variability causes decorrelation which effect produces unsolved direction-dependent errors (DDEs).
2) Faint sources of few milliJanskys of flux that were not cleaned in
the single SBs and that were surrounded by a disk that follows the shape of the PSF. 
This image, with improved resolution, was used to extract a new model that includes 151 sources
\citep[see][]{Yatawatta2013}. 
To take care of the DDEs across the field, we processed the
data with the SAGECAL algorithm \citep{Kazemi2011} by solving in 25 directions
simultaneously, using a hybrid time-solution interval of 2.5 minutes
for directions that included bright sources (flux density $>$ 800 mJy) and 10 minutes for
directions that included faint sources  (flux density $<$ 800 mJy).    
The output of the SAGECAL procedure produced uncalibrated residual
UV-data. Calibration was applied towards the subtracted
sources. Single SBs were imaged with AWimager, stacked, and the model was restored in the final image. 
The resulting image is 5.8\degrees\, wide and has a noise level of 0.8 mJy/beam and
a resolution of 19\arcsec\ $\times$ 18\arcsec. The quality of the image is noticeably
improved but not completely artifact free. However, the distribution of
the residual noise is nearly Gaussian. To improve the image quality and remove
artifacts, another run of self-calibration, using SAGECAL
with the same settings described above, was performed using a model
that includes 1000 sources. 
The final image shown in Fig. \ref{sage2_6deg} covers $5.8 \times
5.8$~deg$^2$ with a resolution of 19\arcsec\ $\times$ 18\arcsec. 
The noise level is 1.3 mJy/beam after the correction of the flux density scale, as described in \ref{fluxes}. 
The catalogue of $\sim$ 1000 sources has been extracted from the final image
by using PyBDSM (version 1.6.1). 
This software decomposes islands of emission into Gaussians, accounting for spatially-varying noise across the image.
The source detection threshold was set to 8 sigma above the mean 
background.
To avoid fitting Gaussians to deconvolution artifacts, PyBDSM
allows the calculation of the rms and mean in a large region over
  the entire image. To better match the typical scale over which the
  artifacts vary significantly, a smaller box near bright sources can
  be defined. In our case, the large box over the entire image was set to 200 
pixels, while a smaller box of 75 pixels was used near sources with
SNR above 50. To  model the extended emission correctly we used the 
wavelet module, which decomposes the residual image that results from the 
fitting of Gaussian into wavelet images at various scales. 

\subsection{Polarization calibration}
\label{polcal}
The images of Q, U, and V Stokes parameters were obtained after the
first BBS run. Imaging  Q and U after SAGECAL would not have been 
appropriate for our purposes, since the time solution interval was
big ($\sim$ 10 min) and the polarized signal could have been suppressed. The U and Q images were used as input for the
RM-Synthesis software.

The WRST model used for the initial calibration is unpolarized, 
which implies that the absolute position angle cannot be determined without observing a polarization calibrator. 

The solutions are independent for each solution interval, which means the polarization vector can rotate by a
different amount each time. This could lead to a loss of
polarized signal when integrating over longer durations of time. However this
does not apply in our case since the solutions were calculated and
applied for each integration time. 

To optimize the prediction of the instrumental polarization,
we solved for the four complex terms of the G-Jones matrix 
and used the theoretical beam model at the same time. Only using  the
beam model prediction would be limited by the accuracy of
the beam model, which is known to be inadequate. 

The estimate of the global TEC time variations for this observation
was done by predicting the amount of ionospheric Faraday rotation and its time
variability \citep{Sotomayor2013}. We found that, through all
the observations, the ionospheric Faraday rotation ranges from values of 
$\sim$ 1.0 to 1.7 rad m$^{-2}$. For wavelengths of 2 m, this corresponds
to Faraday rotation between 4 and 6.8 rad. In such ionospheric
conditions, any polarized signal would be significantly reduced. We
decided not to apply the ionospheric Faraday rotation corrections
since the extreme ionospheric RM variations and the lack of a
polarization calibrator do not allow a quantitative analysis of the
polarized properties of B1834+620.

\subsection{Flux density scale} \label{fluxes} 
Because  the technique of bootstrapping the
calibrator flux scale (see MSSS paper arXiv:1509.01257) was not
implemented in the LOFAR software at the time of observation, the observation was performed without an
observing run on a flux density calibrator. The initial model used for self-calibration was obtained from a WSRT image of the field of B1834+620 at
139\,MHz and resolution of 2\arcmin.  As mentioned in section
\ref{calibration}, we used this model with
BBS to solve  the antenna gains.
However, at the end of the self-calibration process, we discovered that the
flux density scale of this WRST model was about a factor two too low (possibly
caused by the poorly-known beam model of WRST at 2 m wavelength).  
We therefore tied the flux densities of the sources in our image to the extrapolated
flux densities obtained using three well-defined catalogs: VLSS
\citep[][74\,MHz]{VLSS,VLSSr}, WENSS \citep[][325\,MHz]{WENSS}, and NVSS
\citep[][1.4\,GHz]{NVSS}.  Using a matching radius of 20\arcsec\ (according to
the resolution of our image), we selected sources that are detected in our
image in these three catalogs.  We restricted the search to point sources with
a spectral energy distribution that is well-described by a single  power law
($F_\nu\propto \nu^{\alpha}$, with a reduced $\chi^2<2$).  Using this power law, we interpolated the flux density to 144\,MHz ($F_{\rm   ext}$) and
computed the multiplicative factor to apply to the flux density: $F_{\rm ext} /
F_{\rm obs}$.  Restricting the search to sources within two degrees of the
phase center, we obtained 28 matches to the initial WSRT model, yielding a
median flux density ratio (extrapolated over WSRT) of 1.92 with a scatter of
0.11~dex.

We found that the mean of $F_{ext}/F_{obs}$ did not change after applying the
direction-independent calibration; after running BBS, the multiplicative factor
was 1.92 with a scatter of 0.11~dex. However, after the final loop of direction-dependent 
self-calibration using SAGECAL, the factor was 2.11 with a scatter
of 0.09~dex; this corresponds to an uncertainty of $\sim 23\%$ on
  the flux measure. Hence we observed a loss of flux density of about 10\% due to
direction-dependent self-calibration (this effect is usually referred to as self-calibration bias). In the rest of this paper, we  use the corrected flux
densities, i.e., the fluxes measured after the last self-calibration loop,
multiplied by a factor 2.11 (hence the corrected noise level is 1.3 mJy/beam).
The method that we had to apply to normalize our observations implies that the
systematic uncertainty on the flux density of a single source is about
0.1~dex.

\begin{figure*}[]
\begin{center}
\includegraphics[width=18cm]{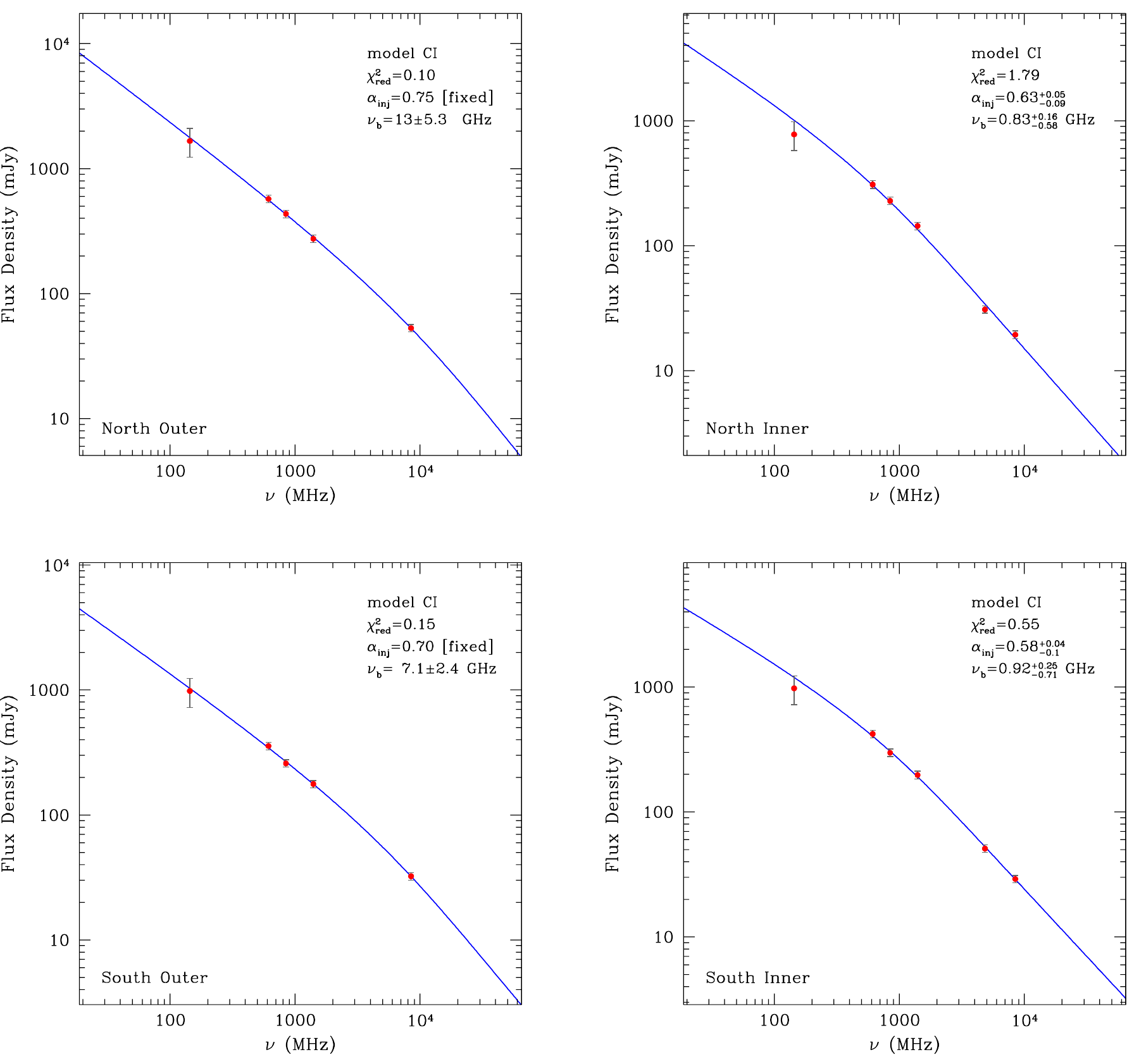}
\end{center}
\caption[]{Radio spectrum of B1834+620. Red dots represent the
  measurements of the flux densities for the four
  components at 144\,MHz (LOFAR, this paper), and 612, 845, 1400, and
8460 \,MHz  \citep[WSRT and VLA][]{Schoenmakers2000a,
  Brocksopp2011}. The blue line indicates the fit with a Continuous Injection model (CI).}
\label{CI}
\end{figure*}

\begin{figure*}[]
\begin{center}
\includegraphics[width=18cm]{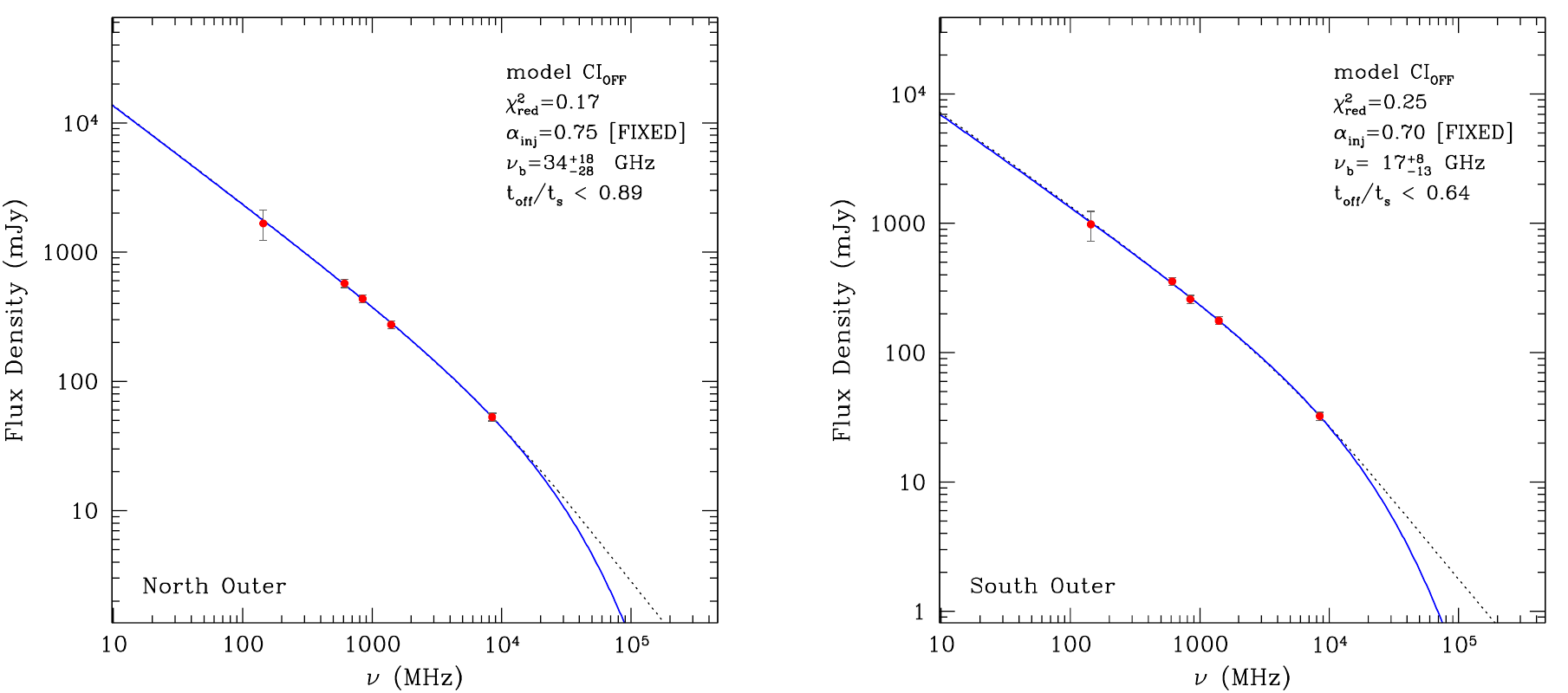}
\end{center}
\caption[]{Radio spectrum of B1834+620, as for fig. \ref{CI}, fitted
  with a Continuous Injection OFF model (CI$_{\rm Off}$).}
\label{CIoff}
\end{figure*}

\begin{figure*}[]
\begin{center}
\includegraphics[width=18cm]{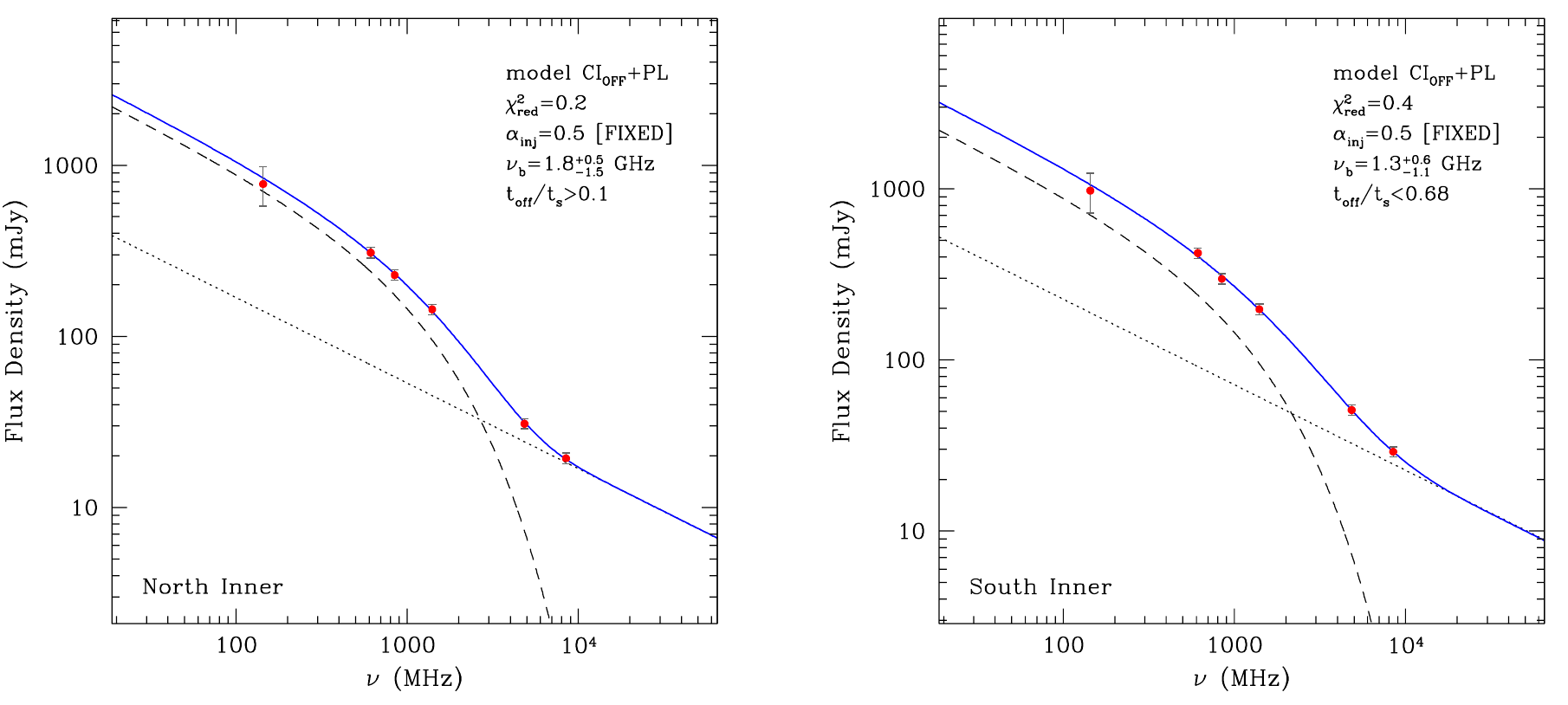}
\end{center}
\caption[]{ Radio spectrum of B1834+620, as for fig. \ref{CI}, fitted
  by a combined model of Continuous Injection OFF (dashed line) and
  power law (dotted line), CI$_{\rm Off}$+PL.}
\label{CIoffPL}
\end{figure*}

\section{Results}

\begin{table}[t]
 \begin{center}
 \caption{Flux densities used in the fits. The measure at 144\,MHz is from our
   LOFAR image, while for the frequencies 612, 845, 1400, 4850, and 8460\,MHz
   we used \cite{Schoenmakers2000a} and \cite{Brocksopp2011}.}
\vspace{5pt}
\medskip
\begin{tabular}{c|cc|cc}
\hline

\noalign{\smallskip}
&   \multicolumn{2}{ c|}{Inner lobe}&\multicolumn{2}{ c}{Outer lobe}\\
&  South & North & South & North\\
\hline

\noalign{\medskip}
$S_{\rm 144\,MHz}$&978$\pm$254&778$\pm$202&984$\pm$256&1666$\pm$433\\
$S_{\rm 612\,MHz}$&422$\pm$30&309$\pm$22 &357$\pm$25&573$\pm$40\\
$S_{\rm 845\,MHz}$&298$\pm$21&228$\pm$16 &259$\pm$18&435$\pm$30\\
$S_{\rm 1400\, MHz}$&198$\pm$14& 144 $\pm$10&177$\pm$12&275$\pm$19\\
$S_{\rm 4850\,MHz}$&51$\pm$3.6&30.9$\pm$2.1& & \\
$S_{\rm 8460\,MHz}$&29.1$\pm$2.0&19.4$\pm$1.4 &32.3$\pm$2.3&53$\pm$ 3.7\\

\noalign{\smallskip}

\hline

\noalign{\smallskip}
\end{tabular}
\end{center}
\end{table}

\begin{table}[t]
  \begin{center}
  \caption{CI fit outer lobes.}

\medskip
  
  \begin{tabular}{ccccc}
\hline
\noalign{\smallskip}
 & \multicolumn{2}{ c| }{South outer lobe}& \multicolumn{2}{ c| }{North outer lobe}\\
    $\alpha_{\rm inj}$ & $\nu_{\rm b}$ &  $\chi^2_{\rm red.}$  &   $\nu_{\rm b}$ &  $\chi^2_{\rm red.}$  \\
            & (GHz)&     &    (GHz) & \\ 
\hline\smallskip\\
0.55  & 1.8$\pm$0.6 & 0.32 &  1.8$\pm$0.7 & 0.49\\
0.60  & 2.8$\pm$1.0 & 0.26 &  2.8$\pm$1.0 & 0.41\\
0.65  & 4.4$\pm$1.5 & 0.19 &  4.5$\pm$1.5 & 0.29\\
0.70  & \bf 7.1$\pm$2.4 & \bf 0.15 & 7.3$\pm$2.5 & 0.16 \\
0.75  & 12.6$\pm$5.0& 0.19 & \bf 13$\pm$5.3 & \bf 0.10\\
0.80  & 27$\pm$11.5 & 0.30 & 28$\pm$13 & 0.11 \\
0.85  & 91$\pm$65&  0.50& 100$\pm$50 & 0.21 \\

\hline
\noalign{\smallskip}
\end{tabular}
\end{center}
\end{table}
\label{summaryfit}

\begin{table}[t]
 \begin{center}
 \caption{CI and CI$_{\rm OFF}$ fits.}
\vspace{5pt}
\medskip
\begin{tabular}{c|cc|cc}
\hline

\noalign{\smallskip}
&   \multicolumn{2}{ c|}{Inner lobe}&\multicolumn{2}{ c}{Outer lobe}\\
&  South & North & South & North\\
\hline

\noalign{\medskip}

\multicolumn{5}{c} {\bf CI model}\\

\noalign{\medskip}
$\alpha_{\rm inj}$&0.58$^{+0.04}_{-0.1}$&0.63$^{+0.05}_{-0.09}$  & 0.70[fixed]& 0.75[fixed]\\
$\nu_{\rm b}$ &0.92$^{+0.25}_{-0.71}$&0.83$^{+0.16}_{-0.58}$ &7.1$\pm$2.4 &13 $\pm$5.3 \\
$\chi^2_{\rm red.}$ &0.55&1.79&0.15&0.10\\

\noalign{\smallskip}

\hline

\noalign{\medskip}

&   \multicolumn{2}{c|}{\bf CI$_{\rm OFF}$+PL} &\multicolumn{2}{c}{\bf CI$_{\rm OFF}$}\\
\noalign{\medskip}
$\alpha_{\rm inj}$&0.5[fixed]&0.5[fixed]& 0.70[fixed]&0.75[fixed] \\
$\nu_{\rm b}$ &1.3$^{+0.6}_{-1.1}$& 1.8$^{+0.5}_{-1.5}$& 17$^{+8}_{-13}$&34$^{+18}_{-28}$  \\
$\chi^2_{\rm red.}$ &0.4&0.2&0.25&0.17\\
$t_{\rm off}/t_{\rm s}$&$<$0.68&$>$0.1&$<0.64$&$<$0.89\\
\noalign{\smallskip}

\hline

\noalign{\smallskip}
\end{tabular}
\end{center}
\end{table}

\subsection{B1834+620}
Figure \ref{B1834+620} shows the LOFAR image of the source B1834+620 at 144\,MHz, obtained
over a bandwidth of between 115.8\,MHz and 162.3\,MHz with a resolution
of 19\arcsec\ $\times$ 18\arcsec. 
The nondetection of the core at this frequency agrees with a convex shape of its spectrum, such as for
gigahertz-peaked sources (GPS) \citep{Schoenmakers2000a,Konar2012}. 

The outer and inner lobes of this DDRG were
resolved, allowing for a comparison between high-
and low-frequency morphologies. 
As shown by \cite{Schoenmakers2000a} and \cite{Konar2012} in
  the image at 8.4\,GHz, in the southern outer lobe there is no
  evidence of a hot spot that confirms its relic nature. The northern outer lobe, on the
  other hand, is characterized by an FR II- type radio morphology.
The high-resolution VLA images at 1.4 and 5\,GHz show that the lobes of the inner double
are FR~II-like, with the inner lobes observed at 1.4\,GHz being
similar to the northern outer lobe at 8.4\,GHz.
The resolution achieved in the LOFAR image at 144\,MHz allowed a more than 10 sigma detection of two new features, elongated from the heads of 
the inner lobes towards the outer lobes. A hint of this emission is
present in 8.4\,GHz and 1.4\,GHz high-resolution images, where a low-brightness bridge, not associated with the inner lobes, appears between the outer and inner
lobes. 
The misalignment of the elongation of the new southern inner lobe feature, with respect to the elongation of the southern
outer lobe, might indicate jet precession \citep{Steenbrugge2008,Falceta2011}. The northern inner lobe presents a brightness elongation
towards the outer lobe that is aligned relative to the direction of the outer lobe's
elongation. Similar features were partly detected at 332.5 MHz by
  \cite{Konar2012} with the Giant Metrewave Radio Telescope (GMRT). 

\begin{figure}[]
\includegraphics[width=8cm]{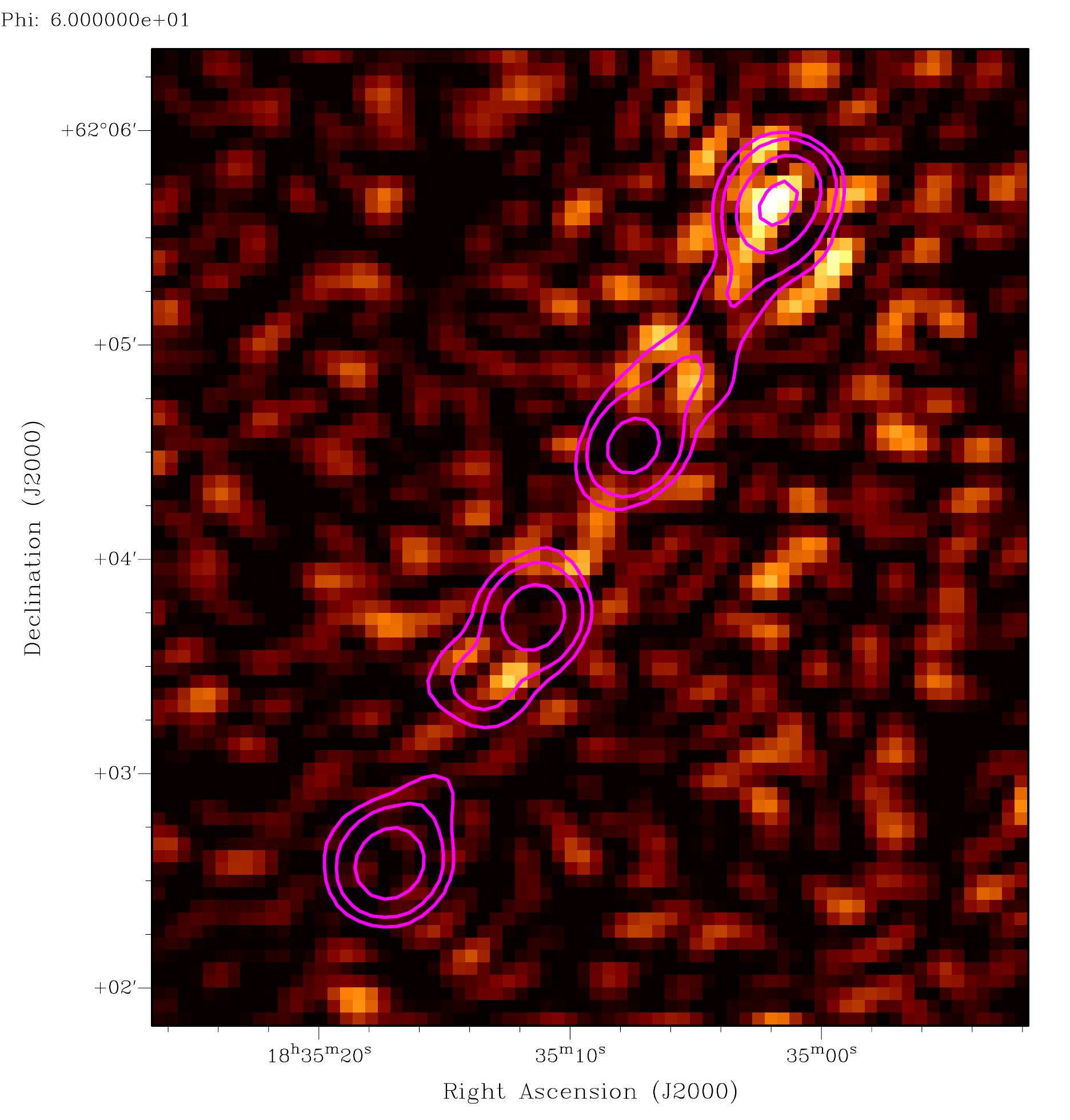}
\includegraphics[width=8cm]{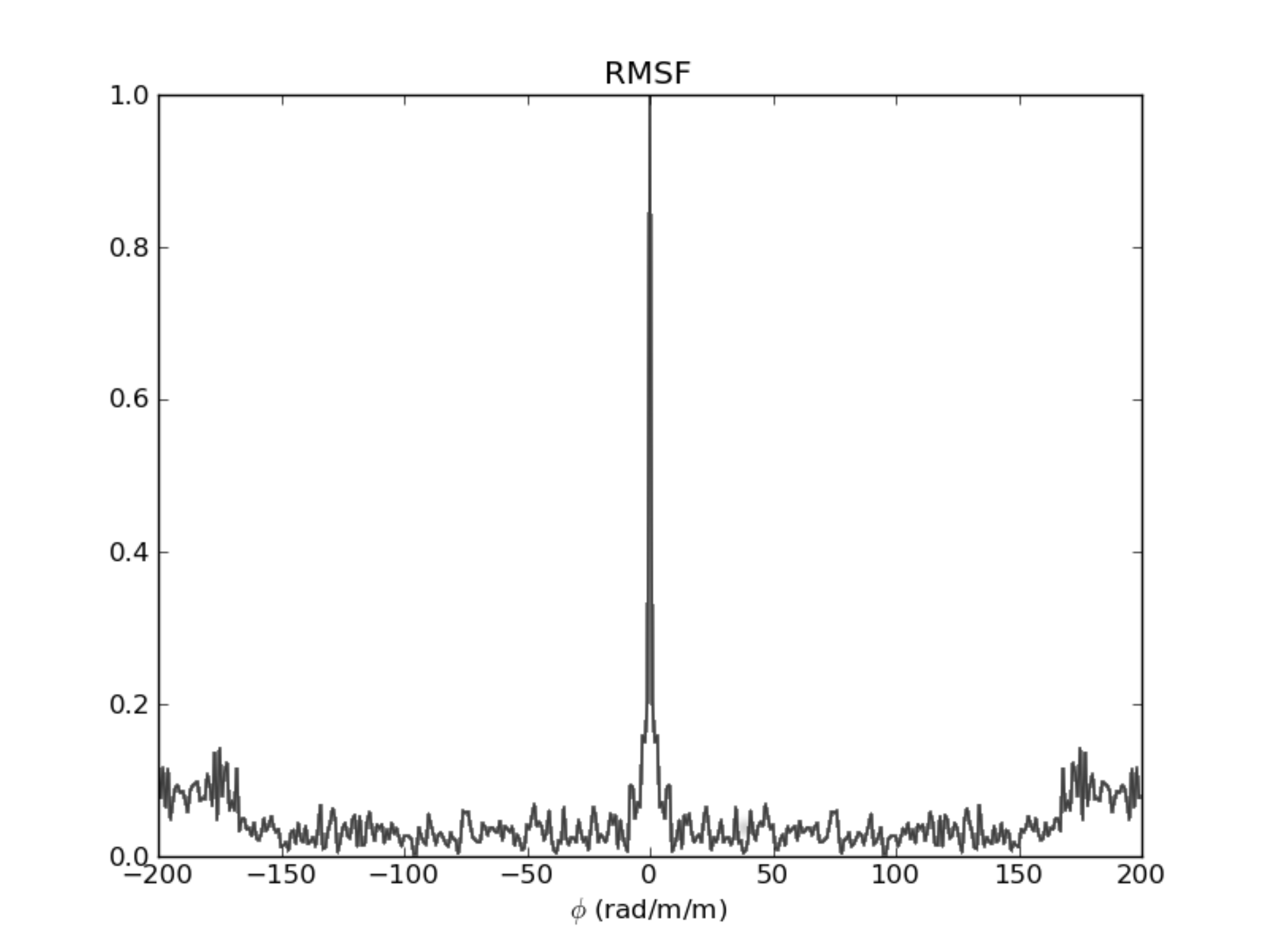}
\includegraphics[width=9cm]{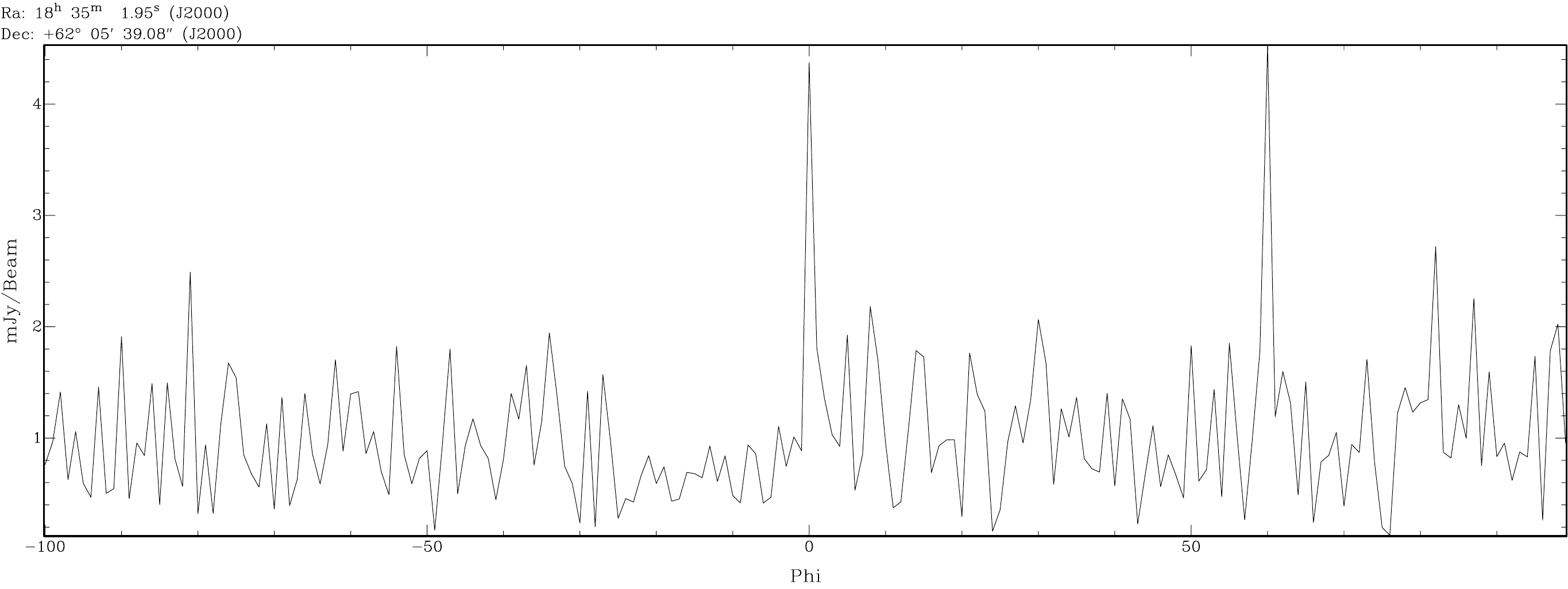}

\caption[]{Top panel: color scale shows the linearly polarized intensity at a
  Faraday depth of +60 rad m$^{-2}$. Middle panel: Response function
  in Faraday depth. Bottom panel: the Faraday spectrum centered on the northern outer lobe. Two peaks are
detected one at  $\phi = +0$ rad m$^{-2}$ (instrumental) and the other at $\phi = +60$ rad m$^{-2}$ (intrinsic). }
\label{pol}
\end{figure} 

The total flux density of 4.7 Jy is in agreement with those found at nearby
  frequencies 
  in the literature  \citep{Schoenmakers2000a,Konar2012} when considering the uncertainty of 23\% already
  discussed earlier in this paper (section \ref{fluxes}).

Red dots in Fig. \ref{CI} show the radio spectrum of the four separate
components; flux density measurements are based on our LOFAR results at 144\,MHz,
together with those reported in \cite{Schoenmakers2000a} and
\cite{Brocksopp2011}, for the frequencies 612, 845, 1400, and
8460\,MHz (Table 1). 
At 144\,MHz we estimate an uncertainty of 23\%  for the flux measurements, based on the scatter obtained in
Section \ref{fluxes}. At high frequency, we assume an error of 7\%, following \cite{Brocksopp2011} . 

To determine the injection spectral index $\alpha_{\rm inj}$
and the break frequency $\nu_{\rm b}$, and to quantify any spectral
ageing of the inner and outer lobes, we fitted the integrated
spectra of the four separate components with a Continuous Injection (CI)
model \citep{Murgia1999}. Since the data we  use are integrated
flux-density measurements, the CI model is the appropriate model to adopt since
the global spectrum of the radio lobes likely originates from a mix 
of electron populations of different ages.
We fitted the spectra of the outer lobes with a CI model spanning a range of possible values of
$\alpha_{\rm inj}$ from 0.55 to 0.85. In Table 2, the resulting fitted
 $\nu_{\rm b}$  parameters are shown for different values of
 $\alpha_{\rm inj}$. We note that the reduced chi-squared values are $<<$ 1.0,
  which implies that the measurement errors are probably
  overestimated. Even if  all the fits were  acceptable formally, we selected the value
  of the fit with the lower $\chi^2_{\rm red.}$. In Fig. \ref{CI} we
  fitted the four components with a CI model. For the outer lobes, we used a fixed
  $\alpha_{\rm inj}$, which corresponds to the best fit shown in bold in
  Table 2. For the inner lobes, $\alpha_{\rm inj}$  is left as free parameter in the
  fit. In both cases, the number of degrees of freedom is three. The results of the fit with the CI model are summarized in the top part of Table 3.

In our best fit we find that both the northern and southern outer lobes are well-fitted with a CI model describing a low-frequency power law with
$\alpha_{\rm inj}$ of 0.75 and 0.70 (Table 2), followed
by a steepening at high frequency with a break at about 13 and 7 GHz for the northern and southern outer lobes respectively.  
Our result agrees within one sigma level with the 
$\alpha_{\rm inj}$=0.8 obtained by \cite{Konar2012} using another
approach: for a different set of frequencies they fitted a
Jaffe-Perola (JP) model \citep{JP} to the two outer lobes together.
The fit of the CI model indicates that  the spectrum at LOFAR frequencies is still the pileup of several
non-aged power laws. The high frequency spectrum is consistent with a
scenario in which the outer lobes are still fed by the central AGN or in which the jets only recently ceased to supply them with
energy, therefore the signatures of electron ageing are not yet
imprinted in the spectral shape. While both scenarios are
  applicable for the northern lobe, this is not true for the southern
 outer lobe. Indeed it shows the characteristics of
  a relic lobe and no hot spot is detected in high
  frequency images at high resolution \citep{Schoenmakers2000a,Konar2012}. To
investigate  the previous scenario, i.e., lobes switched off recently, we fitted the data
with a CI model switched off \citep[CI$_{\rm OFF}$,][]{Murgia2011} with fixed $\alpha_{\rm inj}$ as
found in Table 2. Results are shown in Fig. \ref{CIoff} and summarized in the bottom part of Table 3. In this case the spectra are also well described
by power laws up to high frequencies, 34 and 17 GHz for
the northern and southern lobes, respectively, which is consistent with jets that are still
active or have  recently been switched off. From the fit of the CI$_{\rm OFF}$ model, we
can say that, if the outer lobes are really switched off, the time
spent in the relic phase must be shorter than the one spent in the
active phase, as shown by the fit parameter t$_{\rm off}$/t$_{\rm s}$.  

The fit of the inner lobes with the CI model estimates break frequencies at 830 and 930 MHz, respectively, for the northern and southern inner lobes. 
The $\alpha_{\rm inj}$ is 0.63 and 0.58, respectively, for the northern and
southern inner lobes. The $\alpha_{\rm inj}$ values show no significant
difference  for either the outer or inner lobes, in agreement with \cite{Konar2013a}. 
Despite expectations, the inner lobes, which at the moment should be
actively fed by the central engine, show a relatively low
break frequency compared to the outer
lobes, which are supposed to be relics. 
The $\nu_{\rm b}$ of the outer lobes contradicts the scenario in which 
the outer lobes are the relic lobes and implies that the inner lobes are not the youngest lobes. 
This poses some serious problems, given the morphology of the source. 
We tentatively tried to overcome this problem in the combined fit of Fig. \ref{CIoffPL} by assuming that the
spectrum of the inner double is the mix of two electron populations. The
power law reproduces the emission of the population originating
in the active lobes of the inner doubles (represented by the dotted
lines in Fig. \ref{CIoffPL}) and the CI$_{\rm OFF}$ model 
\citep{Murgia2011} takes into account the emission that possibly
originated in a previous epoch, and of which the newly detected low brightness features
are remnants (represented by the dashed curved lines in
Fig. \ref{CIoffPL}) . The results of the fit with the CI$_{\rm
OFF}$+PL model are summarized in the bottom part of table 3. In this case the estimate of the
relative duration of the dying phase obtained from the fit are not
physically meaningful, since the bow shock of the inner lobes re-accelerate low energy particles that belong to the old jet activity, which  erases their
spectral ageing signatures.  When interpreting the spectral
ageing results, it is important to  note that
re-acceleration mechanism or adiabatic losses could
have played a role in modifying the spectral curvature. Additional spectral data at lower and higher
frequencies would be useful  to model both old and new
emissions.  

The time available for the inner lobes to
be inflated is $\sim$ 2.5 Myrs \citep{Schoenmakers2000a}. 
This is short compared to the time for the outer lobes to expand. 
The standard FR~II model struggles to accommodate the observed
properties of the inner lobes. 
According to the bow shock model proposed by  \cite{Brocksopp2011}, the inner lobes arise from the emission of relativistic electrons
that were compressed and re-accelerated by the bow shock in front of
the jets inside the outer lobes. In this model, the jets in the
inner lobes do not decelerate significantly and the lobes are
expected to expand rapidly, which easily accounts for the relatively
short  lifetime of the inner lobes of B1834+620.
In the  \cite{Brocksopp2011} model, the channels in the existing radio lobes need to collapse before the jet is restarted, otherwise the second jet will  not drive a strong bow shock \citep{Walg14}.
After the bow shocks reach
the tips of the old lobes, the source becomes a standard
FR II-type radio galaxy once more.  
The presence, at low frequency, of two new features related to the
inner lobes of B1834+620 could support the scenario in which the 
jet activity was interrupted for a short time, and the channels of the
existing (outer) lobes are filled with an old electron population that is
compressed and re-accelerated.
Other possible scenarios could interpret the new features as a 
third couple of lobes \citep{Brocksopp2007}.  
As a result of the fit, the $\nu_{\rm b}$ of the inner lobes is 
lower than the one found for the outer lobes. We can  
propose a scenario where the inner lobes are the old lobes and the
outer are the new ones. 

Finally, one of the main results of this paper is the clear detection
of the new features which extend from the tip of the inner double into
the outer lobes. Based on the results shown in Fig. \ref{CIoffPL}), we
could interpret them as the remnant emission of a pair of lobes
that originated during an intermediate burst of the jet activity, which occurred between 
the epochs of the outer and inner lobes. The analysis of the Low Band
Array (LBA) data
(20$-$70 MHz) will allow us to produce a very low-frequency spectral
index distribution that will help us to better separate the old and
new electron populations, and contribute to ruling out or confirming one of
the proposed scenarios, or at least add more pieces to this
complicated puzzle.

\subsection{Polarized emission}
 \cite{Schoenmakers2000a} found that the fractional polarization of
 the inner lobes of B1834+620 is $\sim$20\,$\%,$ both at 1.4\,GHz
 (WSRT$-$VLA) and at 8.4\,GHz Very Large Array (VLA). The same
authors, using VLA and WSRT observations at 612\,MHz, 840\,MHz, 1.4\,GHz, and
8.4\,GHz, constrained the rotation measures (RM) of the four main
components of the source to a range of $\phi$  between +55 and +60 rad m$^{-2}$. 
Polarized emission of about 20$\%$ was observed in the northern outer component of
B1834+620 at 150\,MHz with WSRT. 
The commissioning goal of the observation presented in
this paper was to test the capabilities of the instrument to detect polarized
emission. We produced images of the 
Q, U, and V Stokes parameters after the first BBS run, where an unpolarized model was used.  
Subsequently, we used RM-synthesis \citep{RM2005} to
compute the polarized intensity at Faraday depths between $-$100 and +100 rad m$^{-2}$.  

In the top panel of Fig. \ref{pol}, the color
scale shows the polarized intensity at  $\phi = +60$ rad m$^{-2}$ of the
RM cube; the enhancement of the emission is located in
correspondence with the northern outer lobe. The central panel
  shows the RMSF with a FWHM of $\sim$14 rad m$^{-2}$. The bottom panel presents the
Faraday spectrum centered on the northern outer lobe where two peaks
are detected, one at  $\phi = +0$ rad m$^{-2}$ (where instrumental polarization
ends up) and one at $\phi = +60$ rad m$^{-2}$ (where the intrinsic
polarization measured at $\sim$4$\sigma$ level is concentrated). 
Even though (i) the fractional polarization is lower (P/I=7
  mJy/653 mJy $\sim$\,1\,$\%$) than was found at
higher frequencies by \cite{Schoenmakers2000a}  and by de Bruyn with WSRT and (ii) a reasonable amount of
instrumental polarization is observed at $\phi =  0$ rad m$^{-2}$, our result confirms the value of RM of +60 rad m$^{-2}$
\citep{Schoenmakers2000a}, as found with different radio telescopes and
different methods.  The reason for detecting a lower fractional polarization relative to what is reported as the literature can be identified in the
ionospheric Faraday rotation, which plays a crucial role in depolarizing
the signal.

Despite the severe depolarization caused by the ionospheric Faraday
rotation (up to 6.8 rad, see section \ref{polcal}), the accuracy of the value found
for the RM is larger than the ionospheric variations.
In agreement with \cite{Mulcahy2014}, this observation has proved the capability of LOFAR to detect polarized
emission in AGN.

\subsection{Source counts} 
\label{counts} 
\begin{figure}[] 
\begin{center} 
\includegraphics[trim=4mm 0mm 0mm 6mm, clip,
width=9cm]{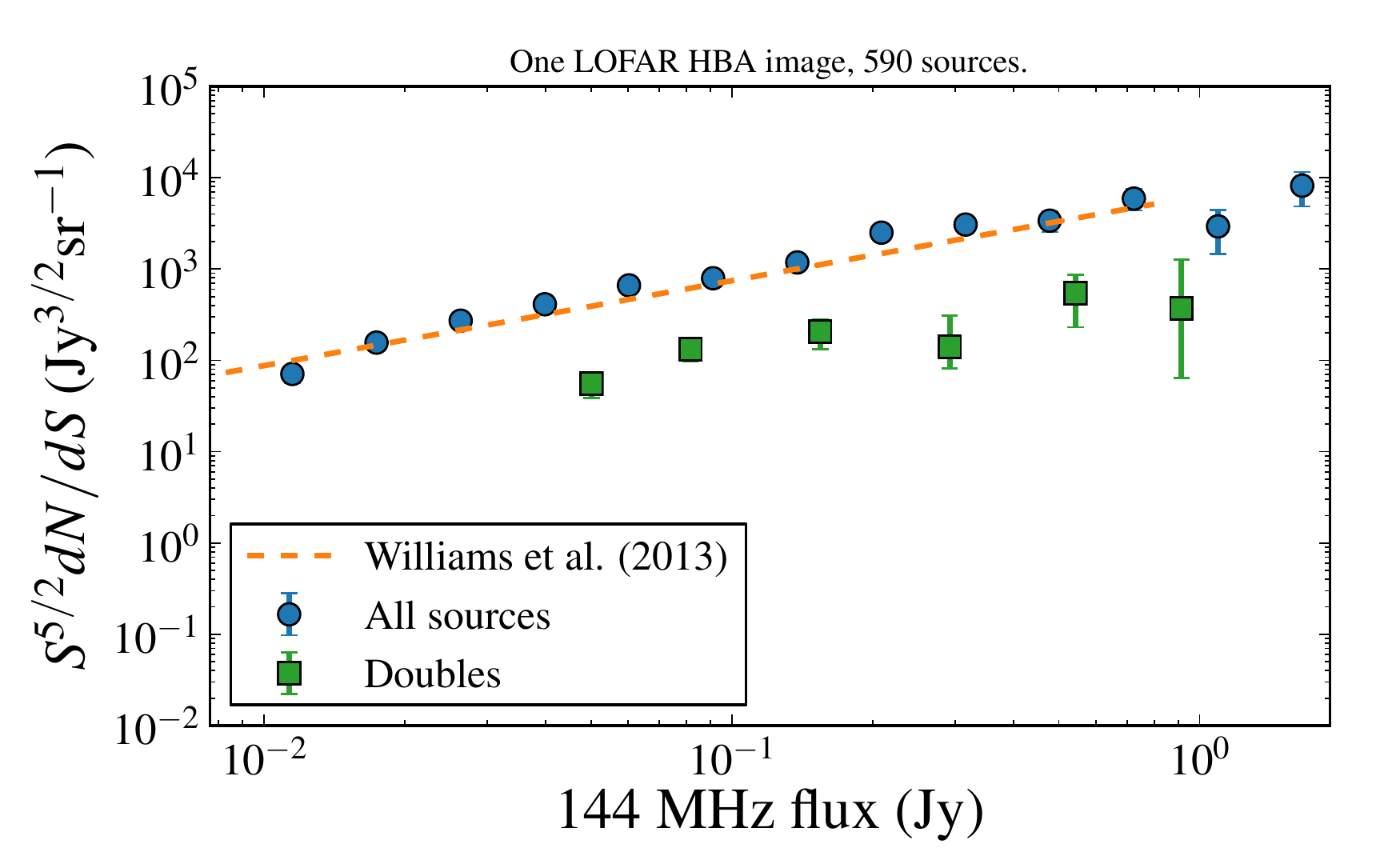} 
\caption{The Euclidian-normalized source counts using a single LOFAR  wide field image. We show the full population and sources with a double-lobed
morphology. For the latter we  subtracted an estimate of the background caused by random matches between two unrelated sources.} \label{dNdS}
\end{center}
\end{figure}

\begin{figure*}[p]
\begin{center}
\includegraphics[]{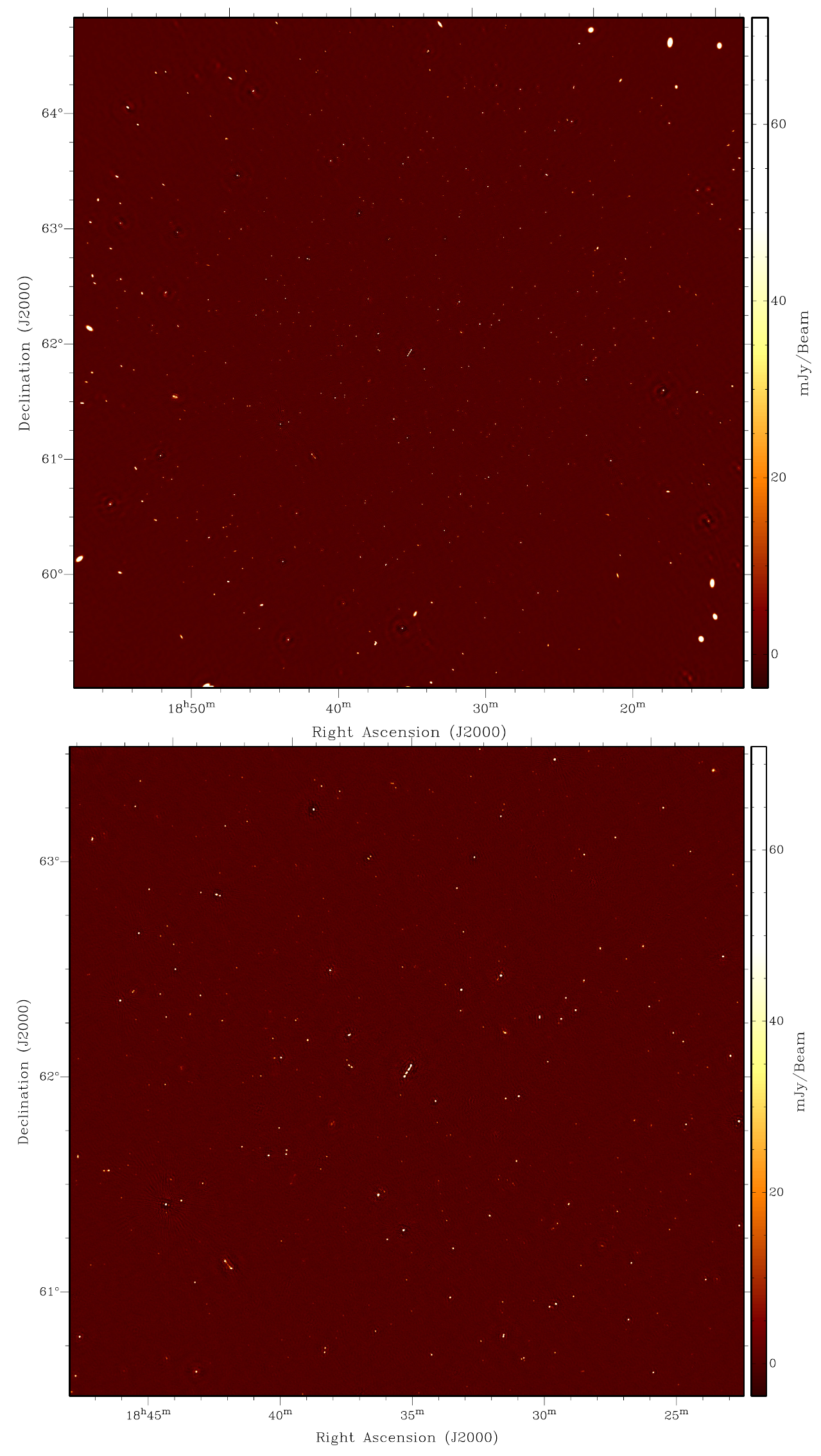}
\end{center}
\caption[]{Color scale shows the intensity of the LOFAR image at 144
  MHz and 19\arcsec\ $\times$ 18\arcsec\ of resolution. Top panel shows a
  field of view of $5.8\times 5.8$~deg$^2$ ($\sim 2$ times the primary
  beam). At the edge of the field, some sources suffer from bandwidth
  smearing. These problems are not seen in the bottom panel, which displays the $2 \times 2$~deg$^2$ area that was used for the source count (section \ref{counts}).}
\label{sage2_6deg}
\end{figure*}

Besides DDRG B1834+620 at the
center of our image, we detected about one thousand more sources, thanks to the
large field of view, imaged as shown in Fig. \ref{sage2_6deg}.  In this section
we present the analysis of the areal density of these sources using the source catalogue produced
by PyBDSM. We restricted the list to the 590 sources located within
2\degrees\, from the phase center (excluding B1834+620). This cut limits
the source extraction to the central part of the image, where the sensitivity is
higher, since the attenuation of the primary beam  is relatively small.  For each
source, we can compute the total area over which it could have been detected,
given our settings of the source-detection algorithm (the rms image
produced by PyBDSM). We computed the differential source count by taking the sum
of the inverse of this area in logarithmically-spaced flux density bins. The result is
shown in Fig.~\ref{dNdS}.  We note that weighting the source count with area
only affects the lowest flux density bin (because the other sources are bright enough to
be detected over the entire image). 
Our finding agrees with the areal density of sources
obtained from multiple GMRT observations of a 30~deg$^2$ region, centered on the
Bootes field \citep{Williams2013}. No deviations from a single power law are
observed, suggesting that a single population (i.e., an AGN) dominates our sample.

\subsection{Doubeltjes} 
A quick visual inspection of the full 34 square degrees
of our image revealed a large number of sources with a double morphology. As
explained in the introduction, we expect  most of these to be FR II AGN at
$z\sim 1$.

To quantify this population of powerful jets, we applied the lobe-finding
algorithm\footnote{The development of this algorithm was, in fact, stimulated
by large number of small doubles that started appearing in high-resolution
LOFAR images.} of \citet[][hereafter VFK15,]{vanVelzen2015} to our source
catalog. We applied this method to the list of Gaussians that have been fitted
to the emission in our image (again, we used only sources found within
2\degrees\, from the center of the field). This algorithm looks for groups
of sources and makes a flux-weighted fit for the symmetry axis of this group.
Often this group is simply a pair. For more complex groups, we remove any
Gaussian that is separated by 20\arcsec\ from the symmetry axis. We identify the
lobes in each group and measure the geometrical center. If a source is found
within 10\arcsec\ of this center we consider this the core of the
FR~II. We excluded sources with lobe-lobe flux ratios greater than a
factor of 3 (which is a slightly more stringent cut compared to what VFK15 used for the FIRST survey).
We also removed sources with lobe-lobe separations larger than 1\arcmin; to take
into account the  resolution of our image, we restrict the selection to pairs with  a
separation that is larger than 20\arcsec\ and a core-to-lobe ratio greater than a
factor 3. These cuts yield 50 doubles with background due to random matches
(as measured using sources drawn from a uniform coordinate
distribution) of 6.2. We manually removed four sources that are misidentified as
doubles caused by imaging artifacts  (mostly bright point sources that are
slightly resolved). We are thus left with 46 candidate FR~IIs, or about 10\% of
the full source list. Because most of these are simply a close pair of point
sources, we sometimes refer to them as doubeltjes (derived from the
Dutch word for `small double'). In Fig.~\ref{dNdS} we show the source count,
after subtracting the background of random matches. 

\begin{figure}[] 
\begin{center} 
\includegraphics[trim=4mm 0mm 0mm 4mm, clip, width=9cm]{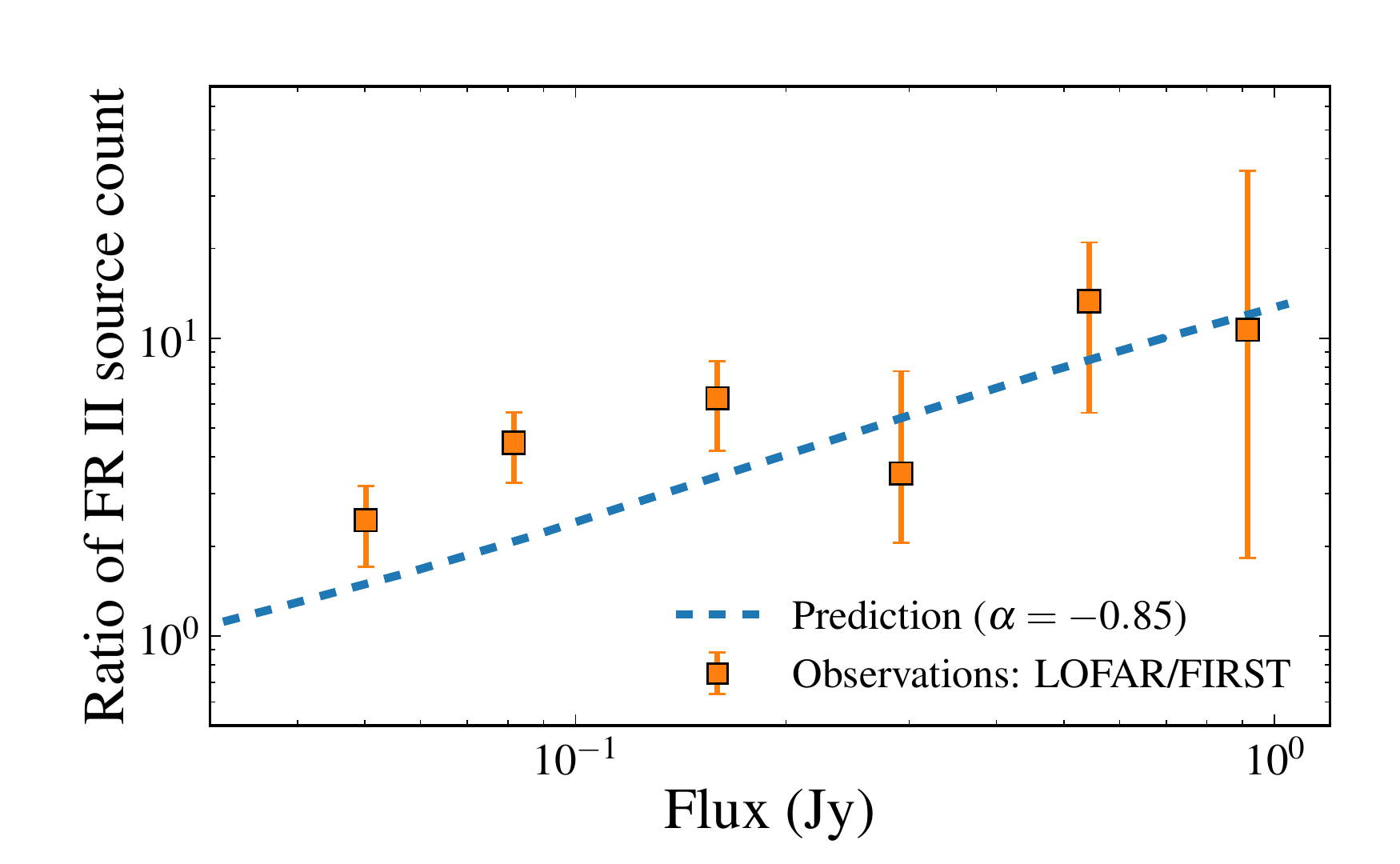} 
\end{center}
\caption{Ratio of the differential source count (number of sources per
flux bin) of double-lobed sources at 144\,MHz and 1.4\,GHz. We show the observed
ratio using our sample of small double-lobed sources from a single LOFAR HBA
image and $\sim 10^4$ `doubeltjes' from the FIRST survey. The expected ratio of
the source count is given by selecting sources from a luminosity function of
jet power and distributing the radio luminosity with a single power law of
index $\alpha=-0.85$. \label{counts_increase}} 
\end{figure}

The areal density of our sample of small double-lobed sources with a total lobe flux of $S_{\rm
144\,MHz} >100$~mJy is 1.6~deg$^{-2}$. Using VFK15 sample, we find that for
$S_{\rm 1.4\,GHz}>100$~mJy,  the areal density of doubeltjes that are selected
using the cuts described above is 0.2~deg$^{-2}$. So, by decreasing the frequency of the
observations from 1.4\,GHz to 144\,MHz, one detects an order of
magnitude that is more doubeltjes. 

The ratio of the differential source count at 144\,MHz and 1.4\,GHz is observed
to increase  with flux (Fig~\ref{counts_increase}).  To understand this
behavior, we first have to understand that, at a given flux density limit, the 144\,MHz
observations probe a source population with a radio luminosity that is lower
by a factor of $10^{-\alpha}$. If the FR~II radio luminosity function flattens at
lower luminosity, a decrease in the low-frequency flux limit  thus results in
a smaller relative increase of the number of detected sources compared to
a decrease in the flux limit at higher frequencies.  Since FR~II morphologies are
typically observed only above a critical radio luminosity
\citep{Fanaroff1974}, a turnover in the FR~II luminosity function is
anticipated. The location of this turnover for an FR~II population at $z\sim
1$, however, is relatively poorly constrained. In the following paragraph, we
present our attempt to reproduce the source count ratio using the luminosity
function of quasars.

To reproduce the ratio of the FR~II source count at two different radio
frequencies (Fig.~\ref{counts_increase}) one needs a model for the evolution
of the FR~II volume density as a function of redshift and luminosity, plus a
model for radio spectral energy distribution of the lobes. Here we used the
measurement of VFK15, who reported the fraction of FR~IIs as a function of the
bolometric luminosity of the accretion disk, $f_{\rm FRII}\propto \exp(-L_{\rm
bol}/L_{\rm 0})$, with $L_0 =10^{46}\, {\rm erg}\,{\rm s}^{-1}$. Following the
methodology of VFK15, we estimate the FR~II areal density by using the
bolometric quasar luminosity function of \citet{Hopkins07} and the optical-radio correlation of FR~II sources \citep[see also][]{vanVelzenFalcke13}. For
each radio-loud AGN, we thus predict a flux density using the observed
correlation between disk luminosity and 1.4\,GHz lobe luminosity. From the
1.4\,GHz luminosity we obtain the 144\,MHz luminosity using the observed
spectral index of the doubeltjes in FIRST, $\alpha=-0.85$ (VFK15).
This yields a prediction for the ratio of the source count that is independent
of the absolute normalization of the luminosity function. Our simple model
reproduces the overall trend, but underestimates the ratio of LOFAR to FIRST
FR~II source count by about 50\% (Fig.~\ref{counts_increase}). This
discrepancy is interesting since it could imply: (i) the number of
FR~IIs from moderate-luminosity quasars ($L_{\rm bol}<10^{45}\,{\rm erg}\,{\rm
s}^{-1}$) is higher than expected; or (ii) a shorter duration of the quasar phase at lower bolometric luminosity, yielding more lobes without
active accretion disks as counterparts; or finally (iii) we are
observing the emergence of a radiatively inefficient AGN population that is
not captured by the \citet{Hopkins07} luminosity function. In the near future,
we will obtain many more doubeltjes from LOFAR fields, allowing us
to study the source count ratio in greater detail.

\section{Conclusions} 
In this paper we presented the first LOFAR image at
144\,MHz of the DDRG B1834+620, in which the four components are resolved.  We
detected emission elongated from the head of the inner lobes towards the outer
lobes,  which is not seen in the higher frequency images of this source. The
radio spectrum between 144\,MHz and 8.5\,GHz was presented. 
The $\alpha_{\rm inj}$ values obtained with the fit
of the CI model, both for the outer and inner lobes, are within the
errors that agree with \cite{Konar2013a}, for which outer
and inner lobes have similar $\alpha_{\rm inj}$ in DDRGs. The spectral fits of the four components are consistent
with the outer lobes, since they are still fed by the central engine or recently
switched off, while the inner lobe spectra is the result of the mix-up of
the emission of new and past jet activity. The presence at low frequency of
two new features related to the inner lobes of B1834+620 seems to support the
model proposed by  \cite{Brocksopp2007}, where these are the channels of the
existing outer  lobes that are filled with an old electron population, but other
scenarios are also possible.  Only the analysis of new LBA LOFAR observations
of this source will provide the spectral index distribution of all the
components that are needed to test these scenarios. This test is important  as
it sheds light on the dynamics of this intriguing radio source and thus helps
us to  understand the AGN duty cycle.  Polarized emission was detected with an RM
of +60 rad m$^{-2}$, confirming the results of \cite{Schoenmakers2000a} and
proving the capability of LOFAR to detect polarized emission from AGN.


Finally, we demonstrated the potential of LOFAR  to detect high-redshift
FR~IIs, using its superior resolution to resolve the two lobes. We  presented
a sample of 46 small doubles separated between 20\arcsec\ and 1\arcmin.  
At a given flux density limit, the areal density of these doubeltjes exceeds the
density of these sources at 1.4\,GHz by a factor of $\sim 10$
(Fig.~\ref{counts_increase}). By extrapolating the observed
source density, we conclude that a 10\,000 square degree LOFAR survey to a
limiting peak flux of 5~mJy,  and at a similar resolution to our image, should
yield $\sim 10^5$ resolved double-lobed radio sources. 

\begin{acknowledgements} 
E.O. would like to thank the anonymous referee for the useful and constructive comments. 
LOFAR, the Low Frequency Array designed and constructed by ASTRON, has facilities in several countries, that are owned by various parties (each with their own funding sources), and that are collectively operated by the International LOFAR Telescope (ILT) foundation under a joint scientific policy.
\end{acknowledgements}

\clearpage
\onecolumn

\bibliographystyle{aa} 
\bibliography{double_orru.bib}

\begin{thebibliography}{51}
\expandafter\ifx\csname natexlab\endcsname\relax\def\natexlab#1{#1}\fi

\bibitem[{{Best}(2009)}]{Best2009}
{Best}, P.~N. 2009, Astronomische Nachrichten, 330, 184

\bibitem[{{Brentjens} \& {de Bruyn}(2005)}]{RM2005}
{Brentjens}, M.~A. \& {de Bruyn}, A.~G. 2005, \aap, 441, 1217

\bibitem[{{Brocksopp} {et~al.}(2007){Brocksopp}, {Kaiser}, {Schoenmakers}, \&
  {de Bruyn}}]{Brocksopp2007}
{Brocksopp}, C., {Kaiser}, C.~R., {Schoenmakers}, A.~P., \& {de Bruyn}, A.~G.
  2007, \mnras, 382, 1019

\bibitem[{{Brocksopp} {et~al.}(2011){Brocksopp}, {Kaiser}, {Schoenmakers}, \&
  {de Bruyn}}]{Brocksopp2011}
{Brocksopp}, C., {Kaiser}, C.~R., {Schoenmakers}, A.~P., \& {de Bruyn}, A.~G.
  2011, \mnras, 410, 484

\bibitem[{{Chandola} {et~al.}(2010){Chandola}, {Saikia}, \&
  {Gupta}}]{Chandola2010}
{Chandola}, Y., {Saikia}, D.~J., \& {Gupta}, N. 2010, \mnras, 403, 269

\bibitem[{{Cohen} {et~al.}(2007){Cohen}, {Lane}, {Cotton}, {Kassim}, {Lazio},
  {Perley}, {Condon}, \& {Erickson}}]{VLSS}
{Cohen}, A.~S., {Lane}, W.~M., {Cotton}, W.~D., {et~al.} 2007, \aj, 134, 1245

\bibitem[{{Condon} {et~al.}(1998){Condon}, {Cotton}, {Greisen}, {Yin},
  {Perley}, {Taylor}, \& {Broderick}}]{NVSS}
{Condon}, J.~J., {Cotton}, W.~D., {Greisen}, E.~W., {et~al.} 1998, \aj, 115,
  1693

\bibitem[{{Di Matteo} {et~al.}(2005){Di Matteo}, {Springel}, \&
  {Hernquist}}]{DiMatteo2005}
{Di Matteo}, T., {Springel}, V., \& {Hernquist}, L. 2005, \nat, 433, 604

\bibitem[{{Fabian} {et~al.}(2006){Fabian}, {Celotti}, \&
  {Erlund}}]{Fabian2006b}
{Fabian}, A.~C., {Celotti}, A., \& {Erlund}, M.~C. 2006, \mnras, 373, L16

\bibitem[{{Falceta-Goncalves} {et~al.}(2011){Falceta-Goncalves}, {Caproni},
  {Abraham}, {de Gouveia Dal Pino}, \& {Teixeira}}]{Falceta2011}
{Falceta-Goncalves}, D., {Caproni}, A., {Abraham}, Z., {de Gouveia Dal Pino},
  E.~M., \& {Teixeira}, D.~M. 2011, ArXiv e-prints [\eprint[arXiv]{1102.0249}]

\bibitem[{{Falcke} {et~al.}(2004){Falcke}, {K{\"o}rding}, \&
  {Markoff}}]{Falcke04}
{Falcke}, H., {K{\"o}rding}, E., \& {Markoff}, S. 2004, \aap, 414, 895

\bibitem[{{Fanaroff} \& {Riley}(1974)}]{Fanaroff1974}
{Fanaroff}, B.~L. \& {Riley}, J.~M. 1974, \mnras, 167, 31P

\bibitem[{{Fender} {et~al.}(2004){Fender}, {Belloni}, \& {Gallo}}]{Fender2004}
{Fender}, R.~P., {Belloni}, T.~M., \& {Gallo}, E. 2004, \mnras, 355, 1105

\bibitem[{{Holt} {et~al.}(2008){Holt}, {Tadhunter}, \& {Morganti}}]{Holt2008}
{Holt}, J., {Tadhunter}, C.~N., \& {Morganti}, R. 2008, \mnras, 387, 639

\bibitem[{{Hopkins} {et~al.}(2007){Hopkins}, {Richards}, \&
  {Hernquist}}]{Hopkins07}
{Hopkins}, P.~F., {Richards}, G.~T., \& {Hernquist}, L. 2007, \apj, 654, 731

\bibitem[{{Jaffe} \& {Perola}(1973)}]{JP}
{Jaffe}, W.~J. \& {Perola}, G.~C. 1973, \aap, 26, 423

\bibitem[{{Kazemi} {et~al.}(2011){Kazemi}, {Yatawatta}, {Zaroubi},
  {Lampropoulos}, {de Bruyn}, {Koopmans}, \& {Noordam}}]{Kazemi2011}
{Kazemi}, S., {Yatawatta}, S., {Zaroubi}, S., {et~al.} 2011, \mnras, 414, 1656

\bibitem[{{Kellermann} {et~al.}(1989){Kellermann}, {Sramek}, {Schmidt},
  {Shaffer}, \& {Green}}]{Kellermann89}
{Kellermann}, K.~I., {Sramek}, R., {Schmidt}, M., {Shaffer}, D.~B., \& {Green},
  R. 1989, \aj, 98, 1195

\bibitem[{{Konar} \& {Hardcastle}(2013)}]{Konar2013a}
{Konar}, C. \& {Hardcastle}, M.~J. 2013, \mnras [\eprint[arXiv]{1309.1401}]

\bibitem[{{Konar} {et~al.}(2012){Konar}, {Hardcastle}, {Jamrozy}, {Croston}, \&
  {Nandi}}]{Konar2012}
{Konar}, C., {Hardcastle}, M.~J., {Jamrozy}, M., {Croston}, J.~H., \& {Nandi},
  S. 2012, \mnras, 424, 1061

\bibitem[{{Konar} {et~al.}(2004){Konar}, {Saikia}, {Ishwara-Chandra}, \&
  {Kulkarni}}]{Konar2004}
{Konar}, C., {Saikia}, D.~J., {Ishwara-Chandra}, C.~H., \& {Kulkarni}, V.~K.
  2004, \mnras, 355, 845

\bibitem[{{K{\"o}rding} {et~al.}(2008){K{\"o}rding}, {Jester}, \&
  {Fender}}]{Koerding2008}
{K{\"o}rding}, E.~G., {Jester}, S., \& {Fender}, R. 2008, \mnras, 383, 277

\bibitem[{{Labiano} {et~al.}(2013){Labiano}, {Garc{\'{\i}}a-Burillo}, {Combes},
  {Usero}, {Soria-Ruiz}, {Tremblay}, {Neri}, {Fuente}, {Morganti}, \&
  {Oosterloo}}]{Labiano2013}
{Labiano}, A., {Garc{\'{\i}}a-Burillo}, S., {Combes}, F., {et~al.} 2013, \aap,
  549, A58

\bibitem[{{Lane} {et~al.}(2014){Lane}, {Cotton}, {van Velzen}, {Clarke},
  {Kassim}, {Helmboldt}, {Lazio}, \& {Cohen}}]{VLSSr}
{Lane}, W.~M., {Cotton}, W.~D., {van Velzen}, S., {et~al.} 2014, \mnras, 440,
  327

\bibitem[{{McNamara} \& {Nulsen}(2007)}]{McNamara2007}
{McNamara}, B.~R. \& {Nulsen}, P.~E.~J. 2007, \araa, 45, 117

\bibitem[{{McNamara} \& {Nulsen}(2012)}]{McNamara2012}
{McNamara}, B.~R. \& {Nulsen}, P.~E.~J. 2012, New Journal of Physics, 14,
  055023

\bibitem[{{Merloni} {et~al.}(2003){Merloni}, {Heinz}, \& {di
  Matteo}}]{Merloni03}
{Merloni}, A., {Heinz}, S., \& {di Matteo}, T. 2003, \mnras, 345, 1057

\bibitem[{{Morganti} {et~al.}(2013){Morganti}, {Fogasy}, {Paragi}, {Oosterloo},
  \& {Orienti}}]{Morganti2013}
{Morganti}, R., {Fogasy}, J., {Paragi}, Z., {Oosterloo}, T., \& {Orienti}, M.
  2013, Science, 341, 1082

\bibitem[{{Morganti} {et~al.}(2005){Morganti}, {Tadhunter}, \&
  {Oosterloo}}]{Morganti2005}
{Morganti}, R., {Tadhunter}, C.~N., \& {Oosterloo}, T.~A. 2005, \aap, 444, L9

\bibitem[{{Mulcahy} {et~al.}(2014){Mulcahy}, {Horneffer}, {Beck}, {Heald},
  {Fletcher}, {Scaife}, {Adebahr}, {Anderson}, {Bonafede}, {Br{\"u}ggen},
  {Brunetti}, {Chy{\.z}y}, {Conway}, {Dettmar}, {En{\ss}lin}, {Haverkorn},
  {Horellou}, {Iacobelli}, {Israel}, {Junklewitz}, {Jurusik}, {K{\"o}hler},
  {Kuniyoshi}, {Orr{\'u}}, {Paladino}, {Pizzo}, {Reich}, \&
  {R{\"o}ttgering}}]{Mulcahy2014}
{Mulcahy}, D.~D., {Horneffer}, A., {Beck}, R., {et~al.} 2014, \aap, 568, A74

\bibitem[{{Murgia} {et~al.}(1999){Murgia}, {Fanti}, {Fanti}, {Gregorini},
  {Klein}, {Mack}, \& {Vigotti}}]{Murgia1999}
{Murgia}, M., {Fanti}, C., {Fanti}, R., {et~al.} 1999, \aap, 345, 769

\bibitem[{{Murgia} {et~al.}(2011){Murgia}, {Parma}, {Mack}, {de Ruiter},
  {Fanti}, {Govoni}, {Tarchi}, {Giacintucci}, \& {Markevitch}}]{Murgia2011}
{Murgia}, M., {Parma}, P., {Mack}, K.-H., {et~al.} 2011, \aap, 526, A148+

\bibitem[{{Offringa} {et~al.}(2010){Offringa}, {de Bruyn}, {Biehl}, {Zaroubi},
  {Bernardi}, \& {Pandey}}]{Offringa2010}
{Offringa}, A.~R., {de Bruyn}, A.~G., {Biehl}, M., {et~al.} 2010, \mnras, 405,
  155

\bibitem[{{Offringa} {et~al.}(2012){Offringa}, {de Bruyn}, \&
  {Zaroubi}}]{Offringa2012}
{Offringa}, A.~R., {de Bruyn}, A.~G., \& {Zaroubi}, S. 2012, \mnras, 422, 563

\bibitem[{{Rengelink} {et~al.}(1997){Rengelink}, {Tang}, {de Bruyn}, {Miley},
  {Bremer}, {Roettgering}, \& {Bremer}}]{WENSS}
{Rengelink}, R.~B., {Tang}, Y., {de Bruyn}, A.~G., {et~al.} 1997, \aaps, 124,
  259

\bibitem[{Romein {et~al.}(2010)Romein, Broekema, Mol, \& van Nieuwpoort}]{CEP}
Romein, J.~W., Broekema, P.~C., Mol, J.~D., \& van Nieuwpoort, R.~V. 2010, in
  ACM Symposium on Principles and Practice of Parallel Programming (PPoPP'10),
  Bangalore, India, 169--178

\bibitem[{{Saikia} \& {Jamrozy}(2009)}]{Saikia2009}
{Saikia}, D.~J. \& {Jamrozy}, M. 2009, Bulletin of the Astronomical Society of
  India, 37, 63

\bibitem[{{Schilizzi} {et~al.}(2001){Schilizzi}, {Tian}, {Conway}, {Nan},
  {Miley}, {Barthel}, {Normandeau}, {Dallacasa}, \& {Gurvits}}]{Schilizzi2001}
{Schilizzi}, R.~T., {Tian}, W.~W., {Conway}, J.~E., {et~al.} 2001, \aap, 368,
  398

\bibitem[{{Schoenmakers} {et~al.}(2000{\natexlab{a}}){Schoenmakers}, {de
  Bruyn}, {R{\"o}ttgering}, \& {van der Laan}}]{Schoenmakers2000a}
{Schoenmakers}, A.~P., {de Bruyn}, A.~G., {R{\"o}ttgering}, H.~J.~A., \& {van
  der Laan}, H. 2000{\natexlab{a}}, \mnras, 315, 395

\bibitem[{{Schoenmakers} {et~al.}(2000{\natexlab{b}}){Schoenmakers}, {de
  Bruyn}, {R{\"o}ttgering}, {van der Laan}, \& {Kaiser}}]{Schoenmakers2000}
{Schoenmakers}, A.~P., {de Bruyn}, A.~G., {R{\"o}ttgering}, H.~J.~A., {van der
  Laan}, H., \& {Kaiser}, C.~R. 2000{\natexlab{b}}, \mnras, 315, 371

\bibitem[{{Sotomayor-Beltran} {et~al.}(2013){Sotomayor-Beltran}, {Sobey},
  {Hessels}, {de Bruyn}, {Noutsos}, {Alexov}, {Anderson}, {Asgekar}, {Avruch},
  {Beck}, {Bell}, {Bell}, {Bentum}, {Bernardi}, {Best}, {Birzan}, {Bonafede},
  {Breitling}, {Broderick}, {Brouw}, {Br{\"u}ggen}, {Ciardi}, {de Gasperin},
  {Dettmar}, {van Duin}, {Duscha}, {Eisl{\"o}ffel}, {Falcke}, {Fallows},
  {Fender}, {Ferrari}, {Frieswijk}, {Garrett}, {Grie{\ss}meier}, {Grit},
  {Gunst}, {Hassall}, {Heald}, {Hoeft}, {Horneffer}, {Iacobelli}, {Juette},
  {Karastergiou}, {Keane}, {Kohler}, {Kramer}, {Kondratiev}, {Koopmans},
  {Kuniyoshi}, {Kuper}, {van Leeuwen}, {Maat}, {Macario}, {Markoff}, {McKean},
  {Mulcahy}, {Munk}, {Orru}, {Paas}, {Pandey-Pommier}, {Pilia}, {Pizzo},
  {Polatidis}, {Reich}, {R{\"o}ttgering}, {Serylak}, {Sluman}, {Stappers},
  {Tagger}, {Tang}, {Tasse}, {ter Veen}, {Vermeulen}, {van Weeren}, {Wijers},
  {Wijnholds}, {Wise}, {Wucknitz}, {Yatawatta}, \& {Zarka}}]{Sotomayor2013}
{Sotomayor-Beltran}, C., {Sobey}, C., {Hessels}, J.~W.~T., {et~al.} 2013, \aap,
  552, A58

\bibitem[{{Spergel} {et~al.}(2003){Spergel}, {Verde}, {Peiris}, {Komatsu},
  {Nolta}, {Bennett}, {Halpern}, {Hinshaw}, {Jarosik}, {Kogut}, {Limon},
  {Meyer}, {Page}, {Tucker}, {Weiland}, {Wollack}, \& {Wright}}]{Spergel2003}
{Spergel}, D.~N., {Verde}, L., {Peiris}, H.~V., {et~al.} 2003, \apjs, 148, 175

\bibitem[{{Steenbrugge} {et~al.}(2008){Steenbrugge}, {Blundell}, \&
  {Duffy}}]{Steenbrugge2008}
{Steenbrugge}, K.~C., {Blundell}, K.~M., \& {Duffy}, P. 2008, \mnras, 388, 1465

\bibitem[{{Tasse} {et~al.}(2013){Tasse}, {van der Tol}, {van Zwieten}, {van
  Diepen}, \& {Bhatnagar}}]{Tasse2013}
{Tasse}, C., {van der Tol}, S., {van Zwieten}, J., {van Diepen}, G., \&
  {Bhatnagar}, S. 2013, \aap, 553, A105

\bibitem[{{van der Tol} {et~al.}(2007){van der Tol}, {Jeffs}, \& {van der
  Veen}}]{Bas2007}
{van der Tol}, S.~., {Jeffs}, B.~D., \& {van der Veen}, A.-J.~. 2007, IEEE
  Transactions on Signal Processing, 55, 4497

\bibitem[{{van Haarlem} {et~al.}(2013){van Haarlem}, {Wise}, {Gunst}, {Heald},
  {McKean}, {Hessels}, {de Bruyn}, {Nijboer}, {Swinbank}, {Fallows},
  {Brentjens}, {Nelles}, {Beck}, {Falcke}, {Fender}, {H{\"o}randel},
  {Koopmans}, {Mann}, {Miley}, {R{\"o}ttgering}, {Stappers}, {Wijers},
  {Zaroubi}, {van den Akker}, {Alexov}, {Anderson}, {Anderson}, {van Ardenne},
  {Arts}, {Asgekar}, {Avruch}, {Batejat}, {B{\"a}hren}, {Bell}, {Bell}, {van
  Bemmel}, {Bennema}, {Bentum}, {Bernardi}, {Best}, {B{\^i}rzan}, {Bonafede},
  {Boonstra}, {Braun}, {Bregman}, {Breitling}, {van de Brink}, {Broderick},
  {Broekema}, {Brouw}, {Br{\"u}ggen}, {Butcher}, {van Cappellen}, {Ciardi},
  {Coenen}, {Conway}, {Coolen}, {Corstanje}, {Damstra}, {Davies}, {Deller},
  {Dettmar}, {van Diepen}, {Dijkstra}, {Donker}, {Doorduin}, {Dromer}, {Drost},
  {van Duin}, {Eisl{\"o}ffel}, {van Enst}, {Ferrari}, {Frieswijk}, {Gankema},
  {Garrett}, {de Gasperin}, {Gerbers}, {de Geus}, {Grie{\ss}meier}, {Grit},
  {Gruppen}, {Hamaker}, {Hassall}, {Hoeft}, {Holties}, {Horneffer}, {van der
  Horst}, {van Houwelingen}, {Huijgen}, {Iacobelli}, {Intema}, {Jackson},
  {Jelic}, {de Jong}, {Juette}, {Kant}, {Karastergiou}, {Koers}, {Kollen},
  {Kondratiev}, {Kooistra}, {Koopman}, {Koster}, {Kuniyoshi}, {Kramer},
  {Kuper}, {Lambropoulos}, {Law}, {van Leeuwen}, {Lemaitre}, {Loose}, {Maat},
  {Macario}, {Markoff}, {Masters}, {McFadden}, {McKay-Bukowski}, {Meijering},
  {Meulman}, {Mevius}, {Middelberg}, {Millenaar}, {Miller-Jones}, {Mohan},
  {Mol}, {Morawietz}, {Morganti}, {Mulcahy}, {Mulder}, {Munk}, {Nieuwenhuis},
  {van Nieuwpoort}, {Noordam}, {Norden}, {Noutsos}, {Offringa}, {Olofsson},
  {Omar}, {Orr{\'u}}, {Overeem}, {Paas}, {Pandey-Pommier}, {Pandey}, {Pizzo},
  {Polatidis}, {Rafferty}, {Rawlings}, {Reich}, {de Reijer}, {Reitsma},
  {Renting}, {Riemers}, {Rol}, {Romein}, {Roosjen}, {Ruiter}, {Scaife}, {van
  der Schaaf}, {Scheers}, {Schellart}, {Schoenmakers}, {Schoonderbeek},
  {Serylak}, {Shulevski}, {Sluman}, {Smirnov}, {Sobey}, {Spreeuw}, {Steinmetz},
  {Sterks}, {Stiepel}, {Stuurwold}, {Tagger}, {Tang}, {Tasse}, {Thomas},
  {Thoudam}, {Toribio}, {van der Tol}, {Usov}, {van Veelen}, {van der Veen},
  {ter Veen}, {Verbiest}, {Vermeulen}, {Vermaas}, {Vocks}, {Vogt}, {de Vos},
  {van der Wal}, {van Weeren}, {Weggemans}, {Weltevrede}, {White}, {Wijnholds},
  {Wilhelmsson}, {Wucknitz}, {Yatawatta}, {Zarka}, {Zensus}, \& {van
  Zwieten}}]{LOFAR2013}
{van Haarlem}, M.~P., {Wise}, M.~W., {Gunst}, A.~W., {et~al.} 2013, \aap, 556,
  A2

\bibitem[{{van Velzen} \& {Falcke}(2013)}]{vanVelzenFalcke13}
{van Velzen}, S. \& {Falcke}, H. 2013, \aap, 557, L7

\bibitem[{{van Velzen} {et~al.}(2015){van Velzen}, {Falcke}, \&
  {K{\"o}rding}}]{vanVelzen2015}
{van Velzen}, S., {Falcke}, H., \& {K{\"o}rding}, E. 2015, \mnras, 446, 2985

\bibitem[{{Walg} {et~al.}(2014){Walg}, {Achterberg}, {Markoff}, {Keppens}, \&
  {Porth}}]{Walg14}
{Walg}, S., {Achterberg}, A., {Markoff}, S., {Keppens}, R., \& {Porth}, O.
  2014, \mnras, 439, 3969

\bibitem[{{Williams} {et~al.}(2013){Williams}, {Intema}, \&
  {R{\"o}ttgering}}]{Williams2013}
{Williams}, W.~L., {Intema}, H.~T., \& {R{\"o}ttgering}, H.~J.~A. 2013, \aap,
  549, A55

\bibitem[{{Yatawatta} {et~al.}(2013){Yatawatta}, {de Bruyn}, {Brentjens},
  {Labropoulos}, {Pandey}, {Kazemi}, {Zaroubi}, {Koopmans}, {Offringa},
  {Jeli{\'c}}, {Martinez Rubi}, {Veligatla}, {Wijnholds}, {Brouw}, {Bernardi},
  {Ciardi}, {Daiboo}, {Harker}, {Mellema}, {Schaye}, {Thomas}, {Vedantham},
  {Chapman}, {Abdalla}, {Alexov}, {Anderson}, {Avruch}, {Batejat}, {Bell},
  {Bell}, {Bentum}, {Best}, {Bonafede}, {Bregman}, {Breitling}, {van de Brink},
  {Broderick}, {Br{\"u}ggen}, {Conway}, {de Gasperin}, {de Geus}, {Duscha},
  {Falcke}, {Fallows}, {Ferrari}, {Frieswijk}, {Garrett}, {Griessmeier},
  {Gunst}, {Hassall}, {Hessels}, {Hoeft}, {Iacobelli}, {Juette},
  {Karastergiou}, {Kondratiev}, {Kramer}, {Kuniyoshi}, {Kuper}, {van Leeuwen},
  {Maat}, {Mann}, {McKean}, {Mevius}, {Mol}, {Munk}, {Nijboer}, {Noordam},
  {Norden}, {Orru}, {Paas}, {Pandey-Pommier}, {Pizzo}, {Polatidis}, {Reich},
  {R{\"o}ttgering}, {Sluman}, {Smirnov}, {Stappers}, {Steinmetz}, {Tagger},
  {Tang}, {Tasse}, {ter Veen}, {Vermeulen}, {van Weeren}, {Wise}, {Wucknitz},
  \& {Zarka}}]{Yatawatta2013}
{Yatawatta}, S., {de Bruyn}, A.~G., {Brentjens}, M.~A., {et~al.} 2013, \aap,
  550, A136

\end{thebibliography}

\end{document}